\documentclass[final]{appolb}
\usepackage{epsfig}

\newcommand{\be}{\begin{equation}}
\newcommand{\ee}{\end{equation}}
\newcommand{\ba}{\begin{eqnarray}}
\newcommand{\ea}{\end{eqnarray}}
\newcommand{\ban}{\begin{eqnarray*}}
\newcommand{\ean}{\end{eqnarray*}}

\begin{document}

\title{Overview of event-by-event fluctuations\thanks{Presented at 
the Fourth Workshop on Particle Correlations and Femtoscopy (WPCF2008), 
Cracow, Poland, September 11-14, 2008}}

\author{Stanis\l aw Mr\' owczy\' nski 
\address{Institute of Physics, Jan Kochanowski University \\
ul.~\'Swi\c etokrzyska 15, PL - 25-406 Kielce, Poland \\
and Andrzej So\l tan Institute for Nuclear Studies \\
ul.~Ho\.za 69, PL - 00-681 Warsaw, Poland}}

\date{February 4, 2009}

\maketitle

\begin{abstract}

Overview of event-by-event studies on relativistic heavy-ion 
collisions is given. I focus on fluctuation measurements and
on theoretical ideas which appeared experimentally fruitful.

\end{abstract}


\section{Introduction}

With the advent of large acceptance detectors it became possible
to observe not one but tens or even hundreds of particles produced 
in a single collision of relativistic nuclei. Such a multi-particle 
state constitutes an {\em event} corresponding to a single 
high-energy collision. Event-by-event analysis is potentially 
a powerful technique to study relativistic heavy-ion collisions,
as magnitude of fluctuations of various quantities around their 
mean values is controlled by system's dynamics. For example, the 
energy and multiplicity fluctuations of many body system are 
related to, respectively, the system's heat capacity and 
compressibility. The two susceptibilities strongly depend the
system's state and they experience dramatic changes at phase 
transitions. So, measuring the fluctuations we can learn about
effective degrees of freedom of the system and their interactions.

Since the early 1990s the event-by-event physics has grown 
to a broad field of active research of relativistic heavy-ion 
collisions, see the review articles 
\cite{Heiselberg:2000fk,Jeon:2003gk}. In the following I overview
the achievements and failures; I discuss difficulties and future 
perspectives of the event-by-event physics. I mostly present 
experimental results and I focus on the theoretical ideas which 
appeared to be experimentally fruitful. Although I have tried to
cover the whole field, the choice of the results to be discussed 
is to some extend subjective.

\section{Early Days Motivation}

The first attractive idea of event-by-event analysis was formulated
by Reinhardt Stock \cite{Stock:1994ve} who suggested to look for 
`interesting' classes of events. The interesting events were meant 
the collisions where the quark-gluon plasma is produced or the 
collisions of exceptionally high multiplicity or energy density 
etc. Imagine there are `hot' events with the temperature 
significantly higher than the average one. Let us further assume 
the event temperature can be quantified by $M(p_T)$ which is
the transverse momentum averaged over particles from a given event. 
It is defined as
\be
M(p_T) \equiv \frac{1}{N} \sum_{i=1}^N p_T^i ,
\ee
where $N$ is the event's multiplicity. If the `hot' events indeed 
exist, then the distribution of $M(p_T)$ should reveal it. 
Fig.~1 shows a typical example of the distribution of $M(p_T)$.
The measurement was performed in central Pb-Pb collisions at 
158 AGeV by NA49 Collaboration at CERN SPS 
\cite{Appelshauser:1999ft}. As seen, no `hot' events are
observed - the $M(p_T)$ distribution  is of boring Gaussian 
shape. 

\begin{figure}[t]
\begin{minipage}{6cm}
\centering
\includegraphics*[width=5.6cm]{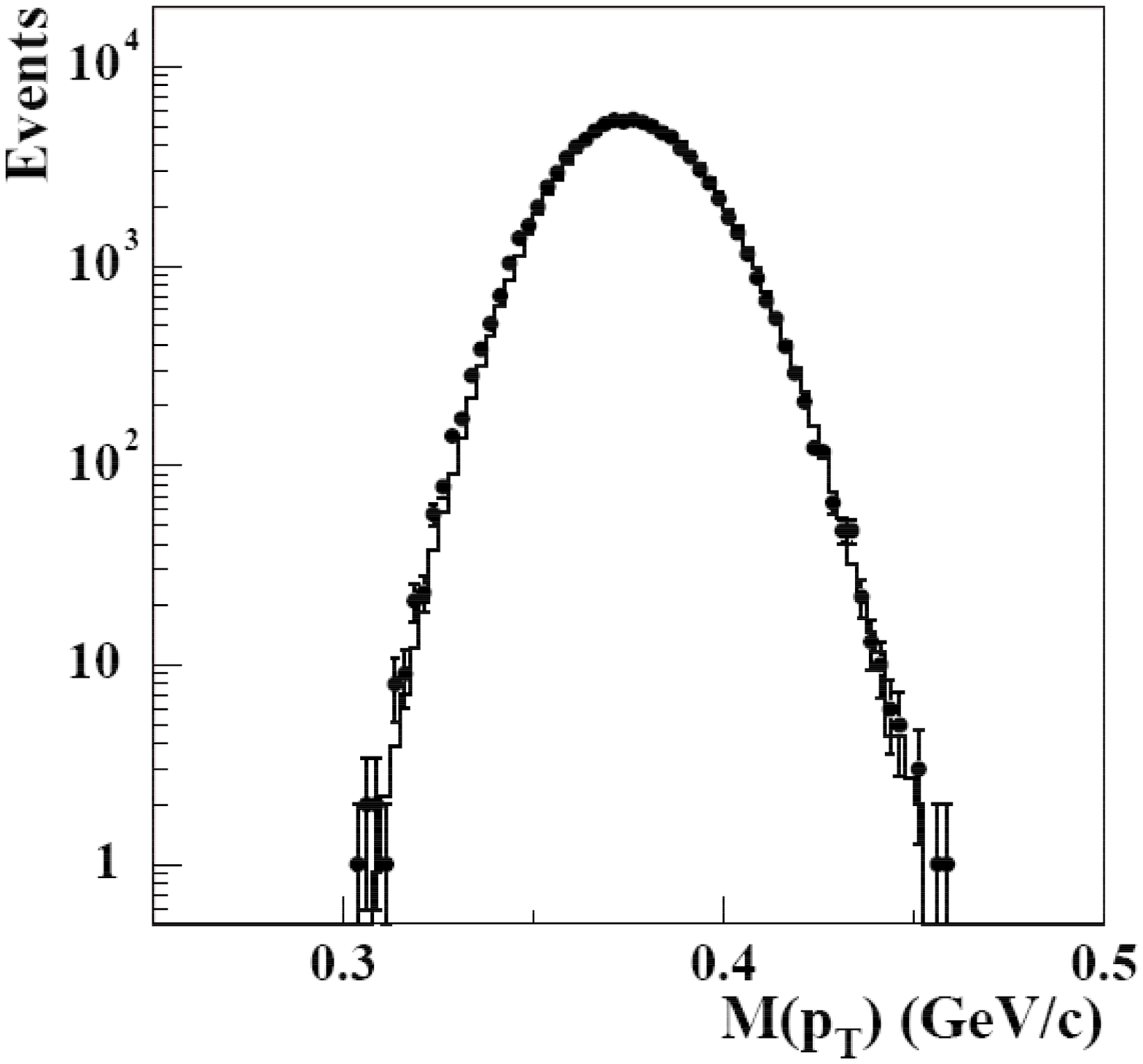}
\caption{The distribution of transverse momentum 
$M(p_T)$ measured in central Pb-Pb collisions at 158 AGeV
by NA49 Collaboration. The histogram and points correspond
to, respectively, the mixed and real events. The figure is 
taken from \cite{Appelshauser:1999ft}}
\end{minipage}
\hspace{2mm}
\begin{minipage}{6cm}
\vspace{-6mm}
\centering
\includegraphics*[width=6cm]{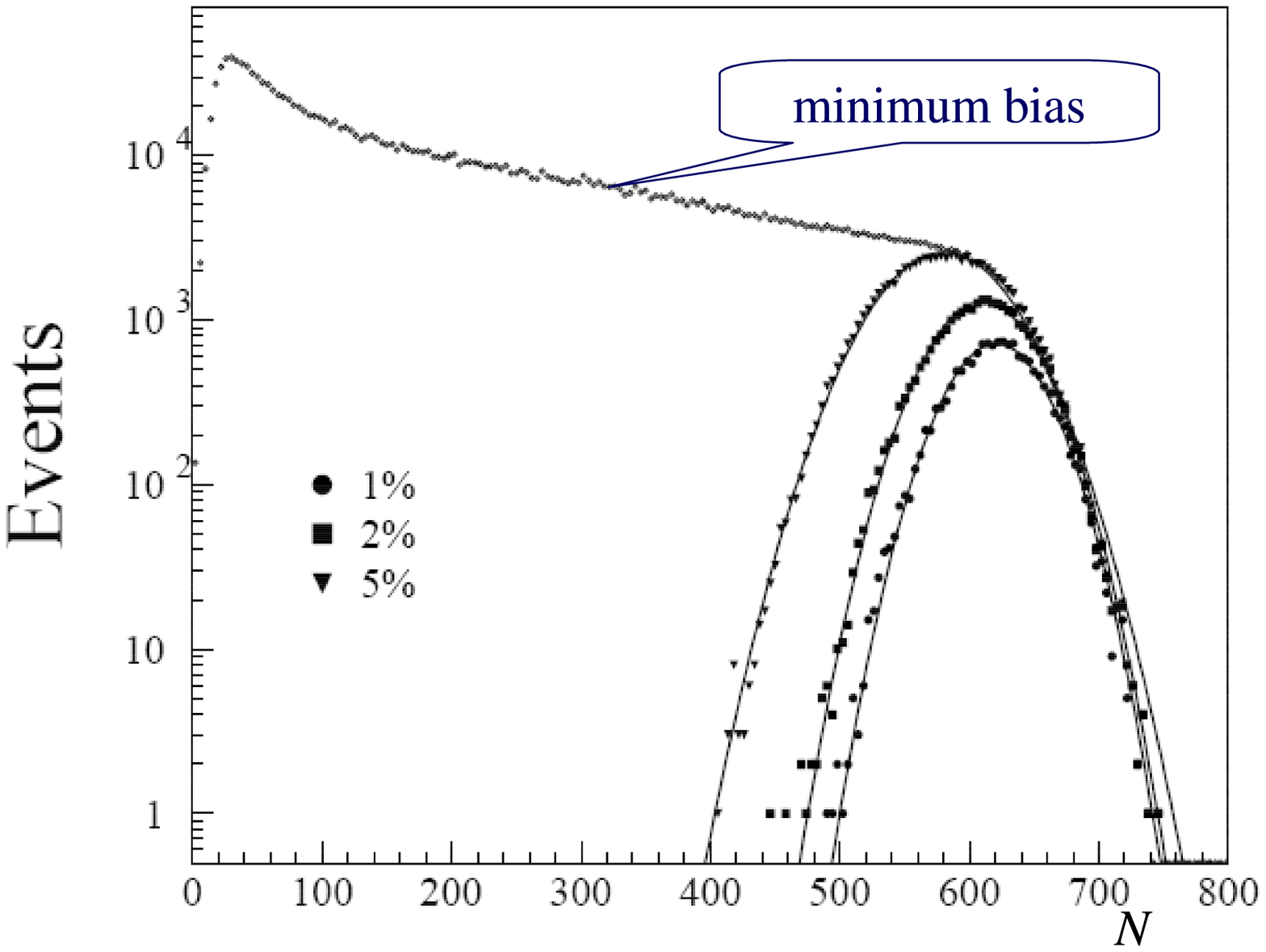}
\vspace{1mm}
\caption{The multiplicity distribution measured in Pb-Pb 
collisions at 158 AGeV by WA98 Collaboration. The minimum
bias data and three classes of central events are shown.
The figure is taken from \cite{Aggarwal:2001aa}}
\end{minipage}
\end{figure}

Fig.~1 shows another typical feature of event-by-event 
distributions. Namely, the distribution of $M(p_T)$ obtained 
for the so-called `mixed' events, where every particle is 
taken from a different event, is nearly identical with that 
obtained for real events. Since there are no inter-particle
correlations in mixed events by construction, the similarity
of the two distributions presented in Fig.~1. shows that 
particles in real events are mostly independent from each 
other. The same is suggested by the Gaussian shape of the 
distribution. The fluctuations present in mixed events
are called {\em statistical} and the fluctuations, which
remain after the statistical fluctuations are subtracted,
are called {\em dynamical}.

Since potentially interesting information encoded in dynamical
fluctuations is not easily seen in the event-by-event 
distributions we have to use more subtle methods to infer it. 
So, in the two next sections I discuss quantities to be measured.
 
\section{Measurable Quantities}
\label{sec-measure-qunat}

\begin{figure}[t]
\vspace{-1mm}
\centering
\includegraphics*[width=9cm]{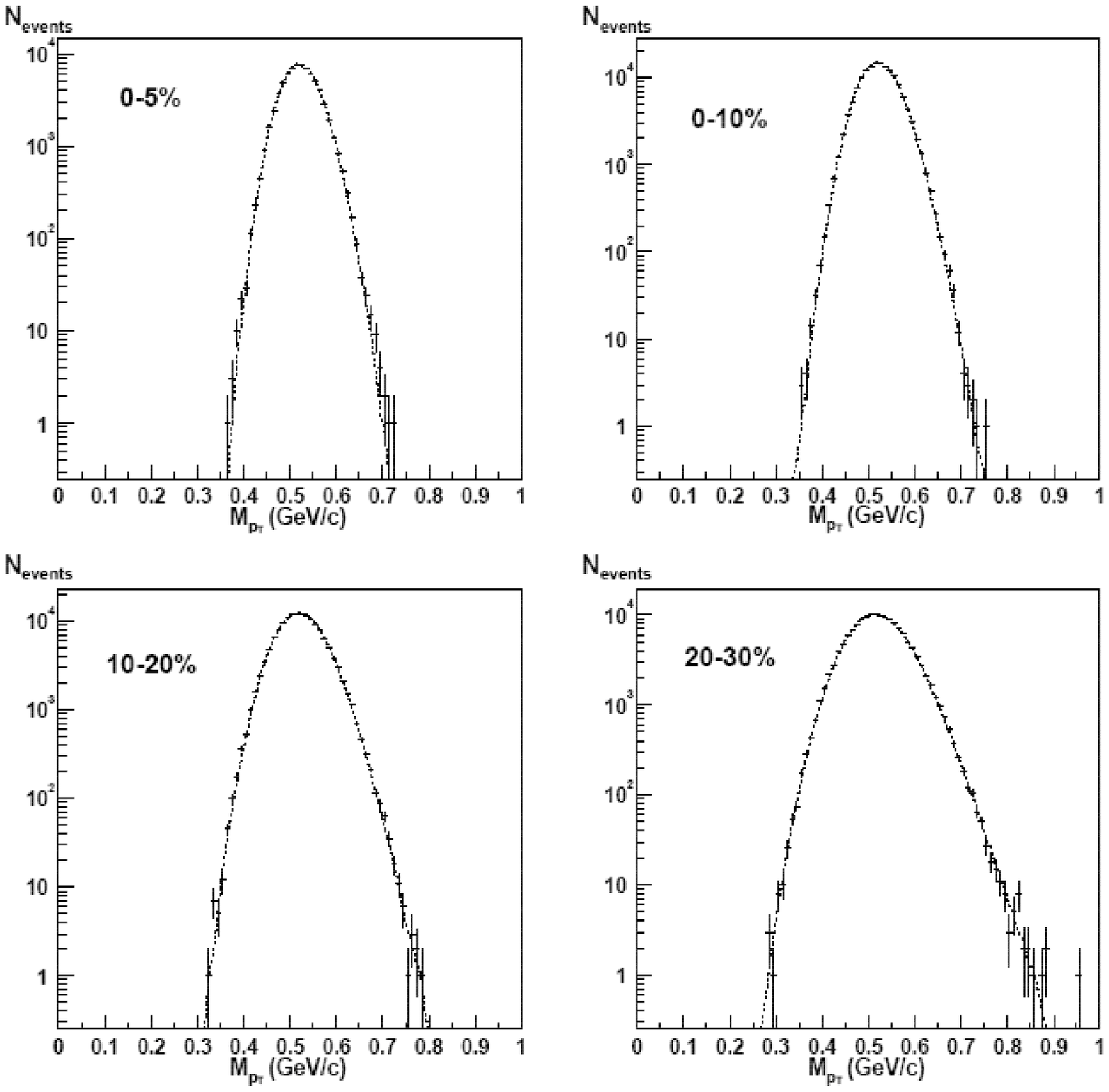}
\caption{The distribution of transverse momentum $M(p_T)$ 
measured in four centrality classes of Au-Au collisions 
at $\sqrt{s_{NN}} = 130 \; {\rm GeV}$ by PHENIX Collaboration
at RHIC. The figure is taken from \cite{Adcox:2002pa}.}
\end{figure}

In thermodynamics we have extensive quantities such as energy
or particle multiplicity, which are proportional to the system's 
volume, and intensive quantities such as temperature or various 
densities, which are independent to the system's size. One is 
tempted to introduce analogous quantities in event-by-event physics
of relativistic heavy-ion collisions. The number of participants
is a natural measure of the size of system which emerges in heavy-ion 
collisions. Then, the quantities like the energy carried by all 
produced particles or particle multiplicity are approximately 
proportional to the number of participants and thus they are
extensive.

Fig.~2 shows the distribution of multiplicity of charged particles
produced in Pb-Pb collisions at 158 AGeV at different centralities.
The measurement was performed by WA98 Collaboration at CERN SPS
\cite{Aggarwal:2001aa}. The collision centrality is defined as 
a percentage of total inelastic cross section $\sigma^{\rm inel}$ 
of nucleus-nucleus collision. The centrality of $n\%$ corresponds, 
roughly speaking, to the collisions with impact parameters from 
such an interval $[0,b]$ that $\pi b^2$ is $n\%$ of 
$\sigma^{\rm inel}$. As seen in Fig.~2, the smaller centrality 
(more central collisions), the higher average multiplicity and 
the smaller width of the distribution. The most upper curve 
corresponds to the minimum bias events when the collisions are
collected with no selection - there is no centrality trigger 
condition. Fig.~2 shows that measurements of extensive quantity 
like multiplicity are not very informative, as the results crucially 
depend on a trigger condition. 

The quantities like $M(p_T)$ are expected to be analogous to 
thermodynamic intensive quantities. Fig.~3 shows the distribution 
of transverse momentum $M(p_T)$ measured in central 
Au-Au collisions at $\sqrt{s_{NN}} = 130 \; {\rm GeV}$ by PHENIX 
Collaboration at RHIC \cite{Adcox:2002pa}. As seen in Fig.~3,
the average value of $M(p_T)$ is indeed approximately independent 
of the system's size (centrality) but the width of the $M(p_T)$
distribution clearly depends on the system's size. And it is unclear
whether the width simply depends on the trigger condition or
it results from dynamics of nuclear collisions.

\section{Fluctuation measures}

In light of previous considerations it is desirable to 
construct a fluctuation measure which is truly intensive 
and it vanishes in absence of inter-particle correlations. 
Several quantities, which satisfy these conditions, have 
been proposed but I focus on the measure $\Phi$ introduced  
in \cite{Gazdzicki:1992ri}. It is constructed as follows. 
One defines the single-particle variable 
$z \equiv x - \overline{x}$ with the overline denoting averaging 
over a single particle inclusive distribution which is performed 
as
\be
\overline{x} = \frac{1}{N_{\rm total}}
\sum_{k=1}^{{\cal N}} \sum_{i=1}^{N_k} x_i
\ee
where $N_k$ is the particle multiplicity in $k-$th event,
${\cal N}$ is the number of events and $N_{\rm total}$ is 
the total number of particles in ${\cal N}$ events. Thus, 
we sum over events and over particles from every event. 
The event variable $Z$, which is a multiparticle analog of $z$, 
is defined as 
\be
Z \equiv
\sum_{i=1}^{N}(x_i - \overline{x}),
\ee 
where the sum runs over particles from a given event. The 
averaging over events is 
\be 
\langle Z \rangle =\frac{1}{\cal N}
\sum_{k=1}^{{\cal N}} Z_k \;.
\ee
One observes that by construction $\langle Z \rangle = 0$. 
Finally, the measure $\Phi$ is defined in the following way
\be
\label{Phi-def}
\Phi \equiv 
\sqrt{\langle Z^2 \rangle \over \langle N \rangle} -
\sqrt{\overline{z^2}} \;.
\ee
The measure $\Phi$ possesses two important properties:

\begin{itemize}

\item
when particles are independent from each other -
there are no correlations among particles coming from
the same event, the $\Phi-$measure vanishes identically;

\item
when particles are emitted by a number of identical
sources, which are independent from each other, $\Phi$
has the same value as for a single source independently
of the distribution of the number of sources ($\Phi$ is
strictly intensive). 

\end{itemize}

Due to the first property $\Phi$ is exactly zero for
mixed events. Because of the second property  
it is strictly independent of centrality in a broad class 
of models of nucleus-nucleus collisions where produced
particles originate form independent sources. The models
include the Wounded Nucleon Model \cite{Bialas:1976ed}
and various models where a nucleus-nucleus collision is 
treated as a superposition of independent nucleon-nucleon 
interactions. In more realistic transport models like
HIJING \cite{Wang:1991hta}, VENUS \cite{Werner:1993uh},
UrQMD \cite{Bass:1998ca} or HSD \cite{Cassing:1999es}, 
there is an admixture of secondary interactions which break 
down independence of nucleon-nucleon interactions. However, 
$\Phi$ is still approximately independent of centrality 
within these models. 

As already mentioned, several other fluctuation measures 
were introduced. In Ref.~\cite{Voloshin:1999yf}, see also 
\cite{Trainor:2000dm}, it was proposed to use 
\be
\label{sigma-dyn}
\sigma^2_{\rm dyn} \equiv 
\langle ( X - \langle X \rangle )^2 \rangle
- \frac{1}{\langle N \rangle} 
\overline{(x- \overline{x})^2} \;,
\ee
where $X$ is the event variable 
\be
X \equiv \frac{1}{N} \sum_{i=1}^{N} x_i .
\ee 
The authors of \cite{Adamova:2003pz} advocated the measure
\be
\Sigma \equiv {\rm sgn}(\sigma^2_{\rm dyn})
\frac{\sqrt{|\sigma^2_{\rm dyn}|}}{\overline{x}} \;.
\ee
We also mention here the quantity $F$ introduced in 
\cite{Adler:2003xq} which is defined in the following way. 
One obtains the scaled dispersion 
\be 
\omega \equiv 
\frac{\sqrt{\langle ( X - \langle X \rangle )^2 \rangle}}
{\langle X \rangle} 
\ee
for real events and for mixed events, and then one computes
\be 
\label{F-def}
F \equiv \frac{\omega_{\rm data} - \omega_{\rm mixed}}
{\omega_{\rm mixed}} \;.
\ee

The fluctuation measures $\sigma^2_{\rm dyn}$, $\Sigma$ and $F$
similarly to $\Phi$ vanish in the absence of inter-particle 
correlations. However, none of these measures is strictly intensive 
as $\Phi$ is. Knowing the average multiplicity $\langle N \rangle$,
the measures $\Phi$, $\sigma^2_{\rm dyn}$, $\Sigma$ and $F$
can be approximately recalculated one into another. 

\section{Transverse Momentum Fluctuations at SPS}
\begin{figure}[t]
\begin{minipage}{6cm}
\centering
\includegraphics*[width=6cm]{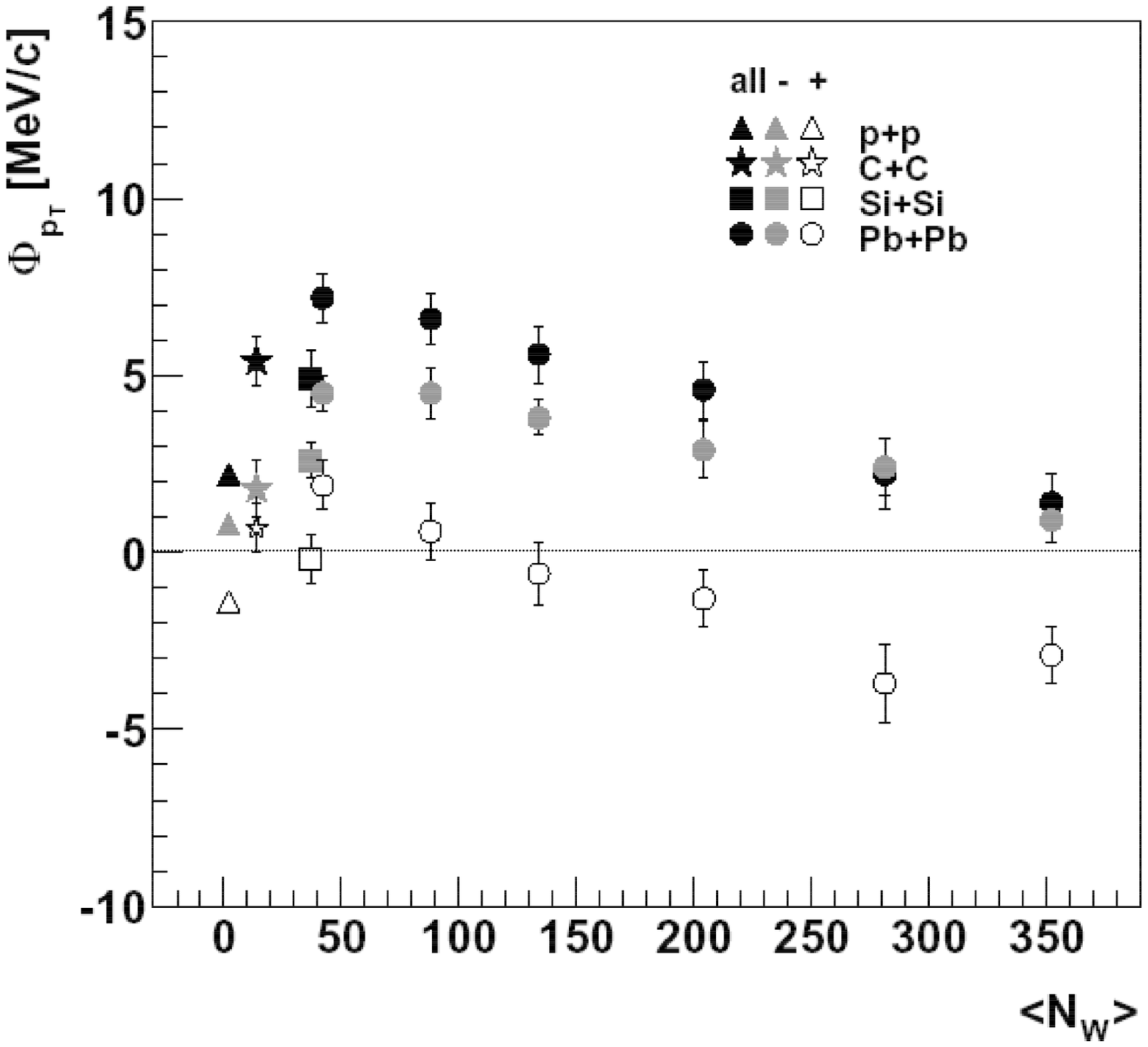}
\caption{$\Phi(p_T)$ as a function of wounded nucleons for
nucleus-nucleus collisions at 158 AGeV. The figure is 
taken from \cite{Anticic:2003fd}}
\end{minipage}
\hspace{3mm}
\begin{minipage}{6cm}
\vspace{-5mm}
\centering
\includegraphics*[width=6cm]{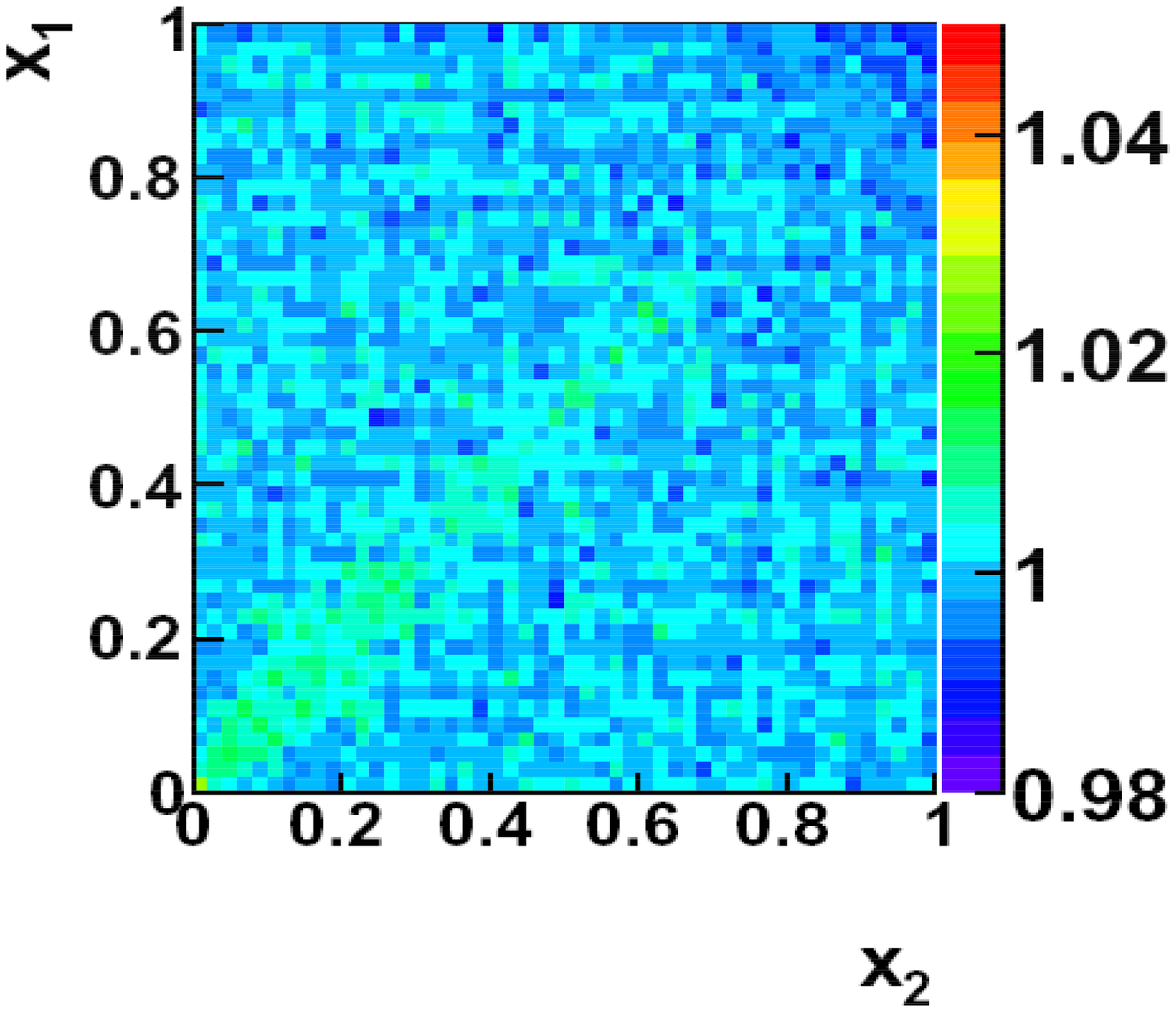}
\caption{Two-particle correlation plot of the cumulant variables 
$x_1$, $x_2$ in central Pb-Pb collisions at 158 AGeV. The figure 
is taken from \cite{Grebieszkow:2007xz}.}
\end{minipage}
\end{figure}

Transverse momentum fluctuations in nucleus-nucleus collisions
at SPS energies were measured by NA49 \cite{Anticic:2003fd}
and CERES \cite{Adamova:2003pz} Collaborations. Fig.~4, which is 
taken from \cite{Anticic:2003fd}, shows the data on p-p, C-C, Si-Si 
and Pb-Pb collisions at 158 AGeV. The fluctuations are measured by 
means of $\Phi$ at various centralities determined by number of
wounded nucleons. In Fig.~1 we can hardly see the difference between 
the real and mixed events, Fig.~4 clearly demonstrates presence of 
dynamical fluctuations and thus it proves sensitivity of the 
$\Phi-$measure. The magnitude of the dynamical correlations 
is quite small ($\Phi \le 8$ MeV) when compared to the dispersion 
of the inclusive transverse momentum distribution (the second term 
in the definition (\ref{Phi-def})) which varies within 
$200 \; {\rm MeV} \le \sigma_{p_T} \le 250 \; {\rm MeV}$ 
\cite{Anticic:2003fd}. So, the dynamical correlation is a few 
percent effect. 

We observe that the fluctuations are different for positive 
and negative particles. It is not surprising as the negative 
particles are nearly all negative pions while the positive particles
include sizeable fraction of protons (the measurement shown
in Fig.~4 was performed in the forward hemisphere). We also
observe the centrality dependence of $\Phi$ with the maximum
at rather peripheral collisions.

Although, the measure $\Phi$ is sensitive to various dynamical
fluctuations, one needs more differential observables to identify
a nature of the fluctuations. For such a purpose one can use
the two-dimensional plot of the cumulant variables $x_1$, $x_2$
proposed in \cite{Trainor:2000dm}. Following \cite{Bialas:1990dk}, 
one defines the cumulat variable 
\be
x(p_T) \equiv \int_0^{p_T}dp_T' P(p_T') .
\ee
where $P(p_T)$ is the inclusive distribution of $p_T$.
Since $P(p_T)$ is normalized to unity, $0 \le x \le 1$.
And now one finds a point $(x_1,x_2)$ for every pair of particles
from the same event and constructs a two-dimensional plot such
as shown in Fig.~5 \cite{Grebieszkow:2007xz}. In the absence
of any correlations the plot is flat and various correlations
generate different patterns in the plot. The example shown in
Fig.~5 proves an existence of positive correlation among particles
of the same $p_T$ which is signaled by the ridge along the diagonal. 
Obviously the correlation is due to the Bose-Einstein statistics
of identical pions.

\begin{figure}[t]
\centering
\includegraphics*[width=12cm]{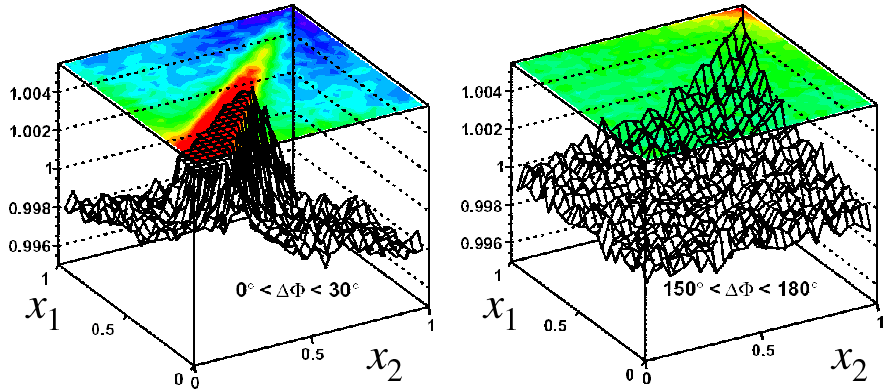}
\caption{Two-particle correlation plot of the cumulant variables 
$x_1$, $x_2$ in central Pb-Au collisions at 158 AGeV. The left 
and right figures, which are taken from \cite{Adamova:2008sx}, 
correspond to the relative azimuthal separation of the two 
particles $0^0 < \Delta \Phi < 30^0$ and 
$150^0 < \Delta \Phi < 180^0$, respectively.}
\end{figure}

The results of even more differential analysis performed by 
CERES Collaboration \cite{Adamova:2008sx} are shown in Fig.~6.
The pairs of particles, which contribute to the correlation 
plot, are divided into classes according to the relative azimuthal 
separation of the two particles $\Delta \Phi$. As seen in Fig.~6,
the pattern of correlation qualitatively changes with $\Delta \Phi$.
For the small separation $0^0 < \Delta \Phi < 30^0$ we observe 
the Bose-Einstein correlation, but for the  maximal separation
$150^0 < \Delta \Phi < 180^0$ the correlation is presumably caused
by the event-by-event fluctuations of the slope of transverse 
momentum distribution. 

The correlation plots shown in Figs.~5, 6 are indeed informative
but still there is a correlation which is not clearly seen in 
these plots. This is the correlation of the event's transverse 
momentum and event's multiplicity which was observed long ago 
in p-p collisions at 205 GeV \cite{Kafka:1976py}. The correlation 
appears to be sufficiently strong to give a significant, if 
not dominant, contribution to $\Phi$ shown in Fig.~4 
\cite{Mrowczynski:2004cg}.

We conclude this section by saying that the dynamical transverse 
momentum fluctuations in heavy-ion collisions at SPS are of various 
physical origin but their total magnitude is quite small.

\section{Transverse Momentum Fluctuations at RHIC}
\label{sec-pT-fluc}

Transverse momentum fluctuations in nucleus-nucleus collisions
at RHIC were measured by PHENIX Collaboration 
\cite{Adler:2003xq} using $F$, see Eq.~(\ref{F-def}), 
and by STAR Collaboration \cite{Adams:2003uw} using 
$\sigma^2_{\rm dyn}$, see Eq.~(\ref{sigma-dyn}). Fig.~7, which 
is taken from \cite{Adler:2003xq}, shows the centrality dependence 
of $p_T$ fluctuations which appears to be similar to 
that at SPS. The magnitude of the fluctuations is bigger. The 
measurement performed by STAR Collaboration \cite{Adams:2003uw}, 
which can be easily recalculated into $\Phi (p_T)$, shows that 
$\Phi (p_T)$ exceeds 50 or even 70 MeV at top RHIC energy. However, 
it is difficult to quantitatively compare results from different 
experiments because the measured fluctuations depend on the 
acceptance which differs from experiment to experiment.

\begin{figure}[t]
\begin{minipage}{6cm}
\centering
\vspace{-7mm}
\includegraphics*[width=6cm]{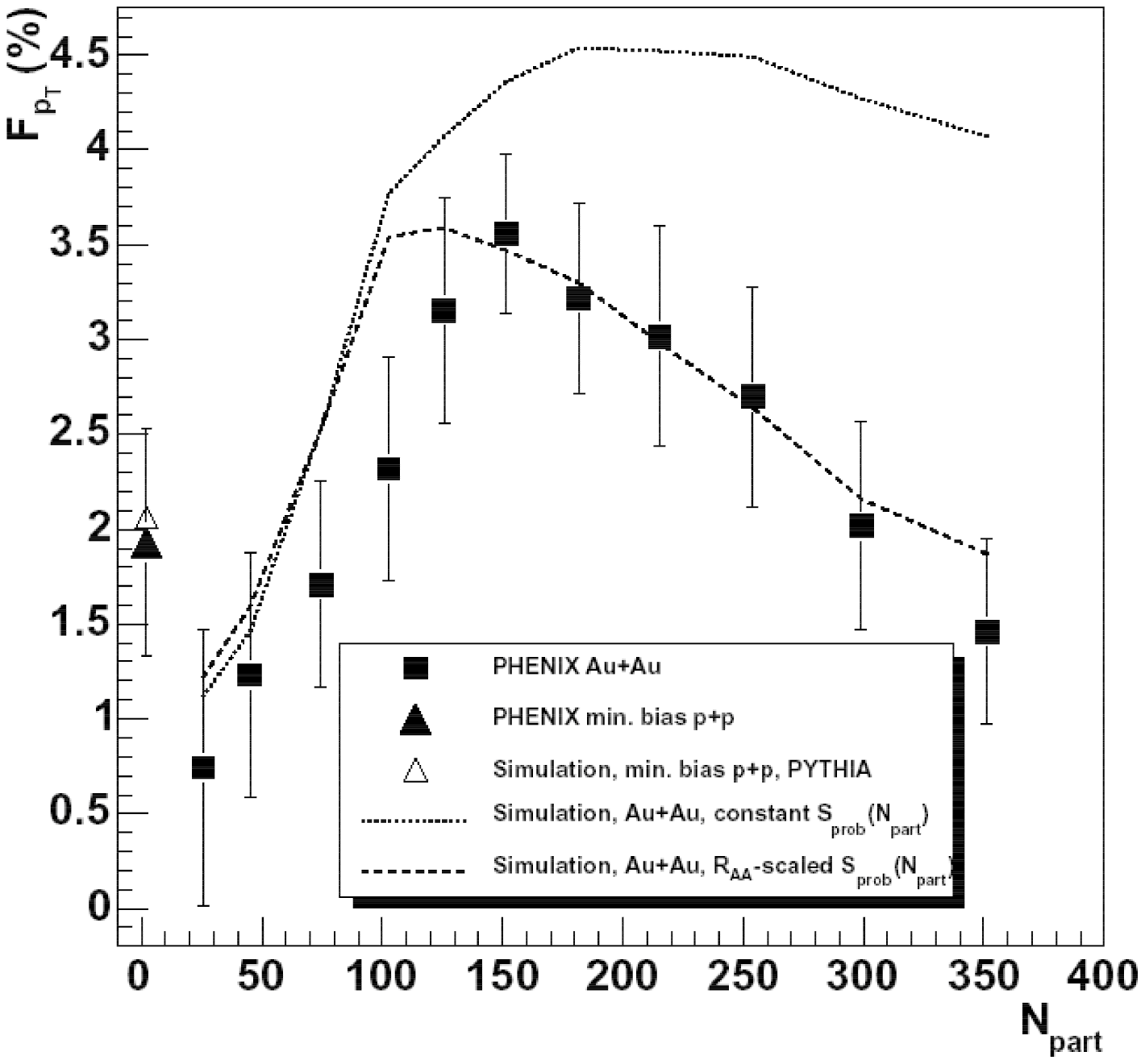}
\caption{$F(p_T)$ as a function of number of participating 
nucleons in Au-Au collisions at $\sqrt{s_{NN}}=200$ GeV. 
The figure is taken from \cite{Adler:2003xq}.}
\end{minipage}
\hspace{3mm}
\begin{minipage}{6cm}
\vspace{-5mm}
\centering
\includegraphics*[width=6cm]{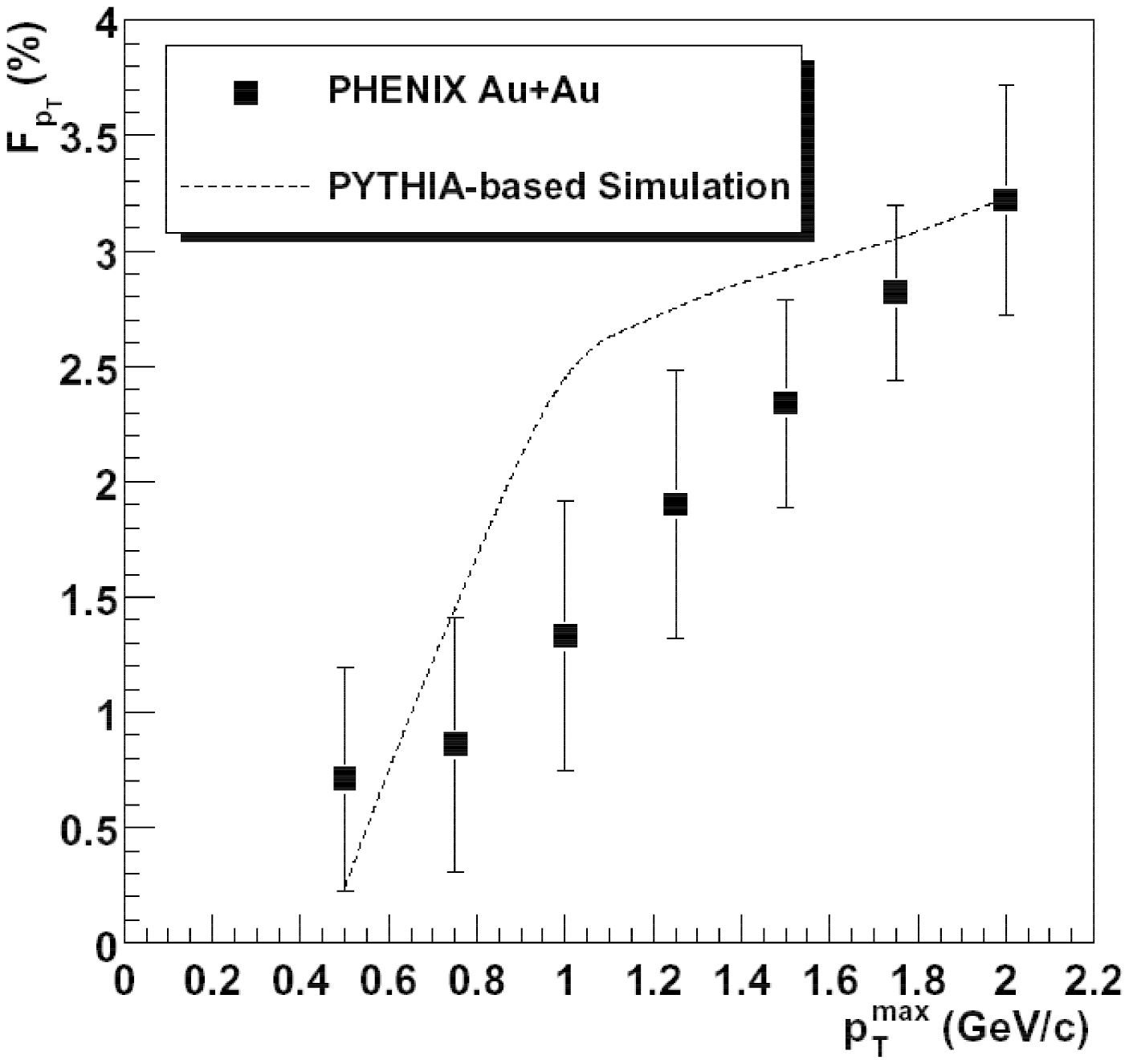}
\caption{$F(p_T)$ as a function of upper $p_T$ cut-off for 
$N_{\rm part} \approx 150$ in Au-Au collisions at 
$\sqrt{s_{NN}}=200$ GeV. The figure is taken from \cite{Adler:2003xq}.}
\end{minipage}
\end{figure}

It was observed in \cite{Adler:2003xq} that the $p_T$ fluctuations
are dominated by particles  with relatively high $p_T$. Fig.~8 shows 
$F(p_T)$ as a function of upper $p_T$ cut-off for the centrality 
corresponding to the maximal fluctuations. For a given $p_T^{\rm max}$ 
only particles with $p_T < p_T^{\rm max}$ are taken into account. 
As seen, $F(p_T)$ grows fast with $p_T^{\rm max}$ and consequently 
it was claimed \cite{Adler:2003xq} that the $p_T$ fluctuations
are due to jets or mini-jets. The claim, however, was questioned 
in \cite{Broniowski:2006zz} where it was argued that the data
from Fig.~8 can be reproduced within a statistical model with
multiple clusters or fireballs which move at some collective 
velocities, correlating the momenta of particles belonging to 
the same cluster. Thus, similarly to the situation at SPS, 
there is no unique interpretation of dynamical $p_T$ fluctuations 
at RHIC.

\section{Thermodynamic fluctuations}

As mentioned in the Introduction, fluctuations in many body 
systems carry information about the system's state and its 
dynamics. Assuming that the strongly interacting matter produced 
in relativistic heavy-ion collisions is in thermodynamic 
equilibrium, it was suggested \cite{Stodolsky:1995ds,Shuryak:1997yj}
to measure the temperature fluctuations. Then, using the relation
\be 
\label{T-fluc}
\langle T^2 \rangle - \langle T \rangle^2 
= \frac{\langle T \rangle^2}{C_V} ,
\ee
which is discussed by Landau and Lifshitz \cite{Lan-Lif}, one can 
infer the system's heat capacity at fixed volume $V$ and particle 
number $N$ 
\be
C_V \equiv \bigg(\frac{\partial U}{\partial T} 
\bigg)_{V,N} \;,
\ee
where $U$ is the system's energy. The relation (\ref{T-fluc}),
however, is actually very controversial \cite{Kit88,Man88} and 
its status is rather unclear. Not entering the details, I think 
that the relation (\ref{T-fluc}) cannot be used, as long as the 
thermometer to measure the temperature fluctuations is not 
specified \cite{Stephanov:1999zu}.
 
A similar idea \cite{Mrowczynski:1997kz} was to infer the 
compressibility 
\be
\kappa \equiv -\bigg(\frac{\partial p}{\partial V} 
\bigg)_{T,\langle N \rangle} \;,
\ee
where $p$ is the pressure, from the multiplicity 
fluctuations due to the relation \cite{Lan-Lif}
\be 
\label{N-fluc}
\langle N^2 \rangle - \langle N \rangle^2 
= \frac{T \langle N \rangle^2}{V^2 \kappa} .
\ee
An experimental problem here is to measure the multiplicity 
fluctuations at fixed system's volume. 

Only the third idea to study electric charge fluctuations in
relativistic heavy-ion collisions appeared to be experimentally
relevant. The fluctuations are related to the electric charge 
susceptibility \cite{Jeon:2003gk} as
\be 
\label{Q-fluc}
\langle Q^2 \rangle - \langle Q \rangle^2 
= TV \chi_Q ,
\ee
with
\be
\label{Q-suspect}
\chi_Q \equiv - \bigg(\frac{\partial F}{\partial \mu_Q} 
\bigg)_{T,V} \;,
\ee
where $F$ is the free energy and $\mu_Q$ is the chemical
potential responsible for the electric charge conservation.
Eqs.~(\ref{Q-fluc}, \ref{Q-suspect}) do not look very 
exciting at first glance but it was sharply observed 
\cite{Jeon:2000wg,Asakawa:2000wh} that the susceptibility 
(\ref{Q-suspect}) is very different in the quark phase and 
in the hadron one.

To explain this statement, let me consider the classical ideal 
gas of particles of chargers $\pm  q$ (measured in the units of
elementary charge). The system's charge is then $Q = q(N_+-N_-)$. 
We introduce $\delta Q \equiv Q - \langle Q \rangle$ and 
$\delta N_{\pm} \equiv N_{\pm} - \langle N_{\pm} \rangle$
and we compute the charge fluctuations as
$$
\langle (\delta Q)^2 \rangle = 
q^2 \langle (\delta N_+- \delta N_-)^2 \rangle =
q^2 \Big(\langle (\delta N_+)^2 \rangle 
+ \langle \delta N_-)^2 \rangle - 
2 \langle \delta N_+ \delta N_-\rangle \Big) .
$$
Since in the ideal classical gas 
$\langle (\delta N_\pm)^2 \rangle = \langle N_\pm \rangle$
and $\langle \delta N_+ \delta N_-\rangle = 0$, one finds
\be
\label{Q-fluc-per-part}
\frac{\langle (\delta Q)^2 \rangle}{\langle N \rangle} = q^2 .
\ee
where $\langle N \rangle \equiv 
\langle N_+ \rangle + \langle N_- \rangle$. As seen in 
Eq.~(\ref{Q-fluc-per-part}), the charge fluctuation per particle 
equals the particle's charge squared.

One easily derives the formula analogous to 
Eq.~(\ref{Q-fluc-per-part}) for the ideal classical gas 
of pions composed of $\pi^+,\;\pi^-,\;\pi^0$ and for 
the quark-gluon plasma being a mixture of ideal classical
gases of quarks of different charges and of neutral gluons.
Using the system's entropy $S$ instead of the total particle 
multiplicity $\langle N \rangle$, one finds \cite{Jeon:2003gk}
\be
\label{Q-fluc-per-entropy}
\frac{\langle (\delta Q)^2 \rangle}{S} =
\left\{
\begin{array}{cc}
\frac{1}{6} & {\rm for \; pions,} \\[2mm]
\frac{1}{24} & {\rm for \; QGP.} 
\end{array}
\right.
\ee
It was argued in \cite{Jeon:2000wg,Asakawa:2000wh} that
the charge fluctuations generated in the quark phase
are frozen due the system's fast hydrodynamic expansion 
and that the entropy, which is mostly produced at the 
very early, preequilibrium  stage of the collision, is 
approximately conserved during the hydrodynamic evolution. 
Then, a measurement of the ratio (\ref{Q-fluc-per-entropy}) 
should clearly show whether the quark-gluon plasma is produced 
at the early stage of relativistic heavy-ion collisions.

\begin{figure}[t]
\begin{minipage}{6cm}
\centering
\vspace{-7mm}
\includegraphics*[width=6cm]{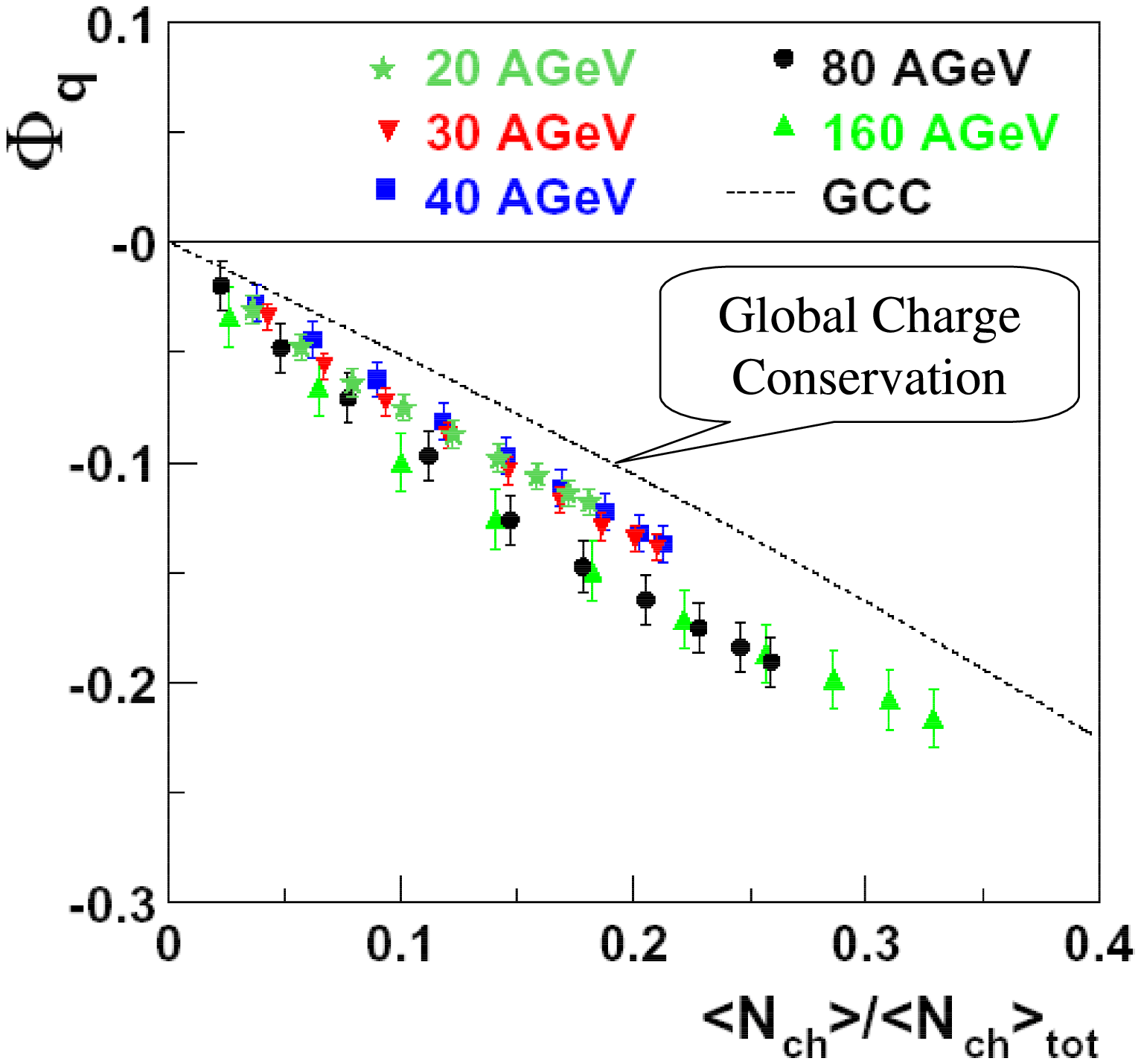}
\caption{Electric charge fluctuations quantified by $\Phi_q$
as a function of relative charge multiplicity in central
Pb-Pb collisions at SPS for several collision energies. 
The figure is taken from \cite{Alt:2004ir}}
\end{minipage}
\hspace{3mm}
\begin{minipage}{6cm}
\vspace{-5mm}
\centering
\includegraphics*[width=6cm]{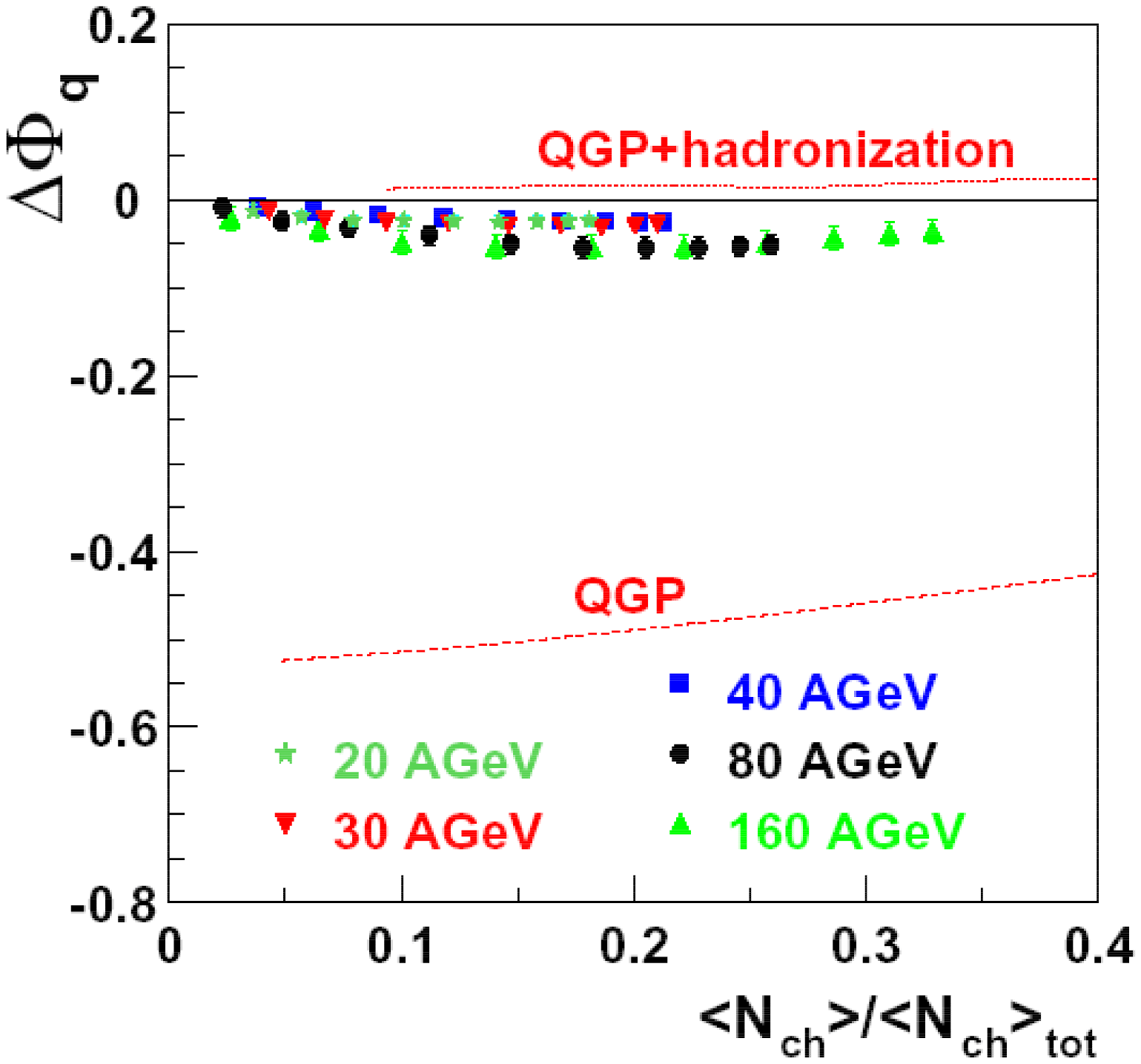}
\caption{Electric charge fluctuations quantified by 
$\Delta\Phi_q \equiv \Phi_q -\Phi_q^{\rm GCC}$ as a function 
of relative charge multiplicity in central Pb-Pb collisions 
at SPS for several collision energies. The figure is taken 
from \cite{Alt:2004ir}.}
\end{minipage}
\end{figure}

Soon later the electric charge fluctuation were measured
experimentally. Figs. 9 and 10 show the results obtained
at SPS by NA49 Collaboration \cite{Alt:2004ir} using the 
measure $\Phi$ defined by Eq.~(\ref{Phi-def}). 
$\langle N_{\rm ch} \rangle_{\rm tot}$ and 
$\langle N_{\rm ch} \rangle$ are the average charge 
particle multiplicities in, respectively, the full ($4\pi$) 
acceptance and in a given phase-space domain under study.
As seen, the results, which are essentially independent of 
the collision energy, follow the trend dictated by the 
global charge conservation (GCC) corresponding to 
\be
\label{GCC}
\Phi_q^{\rm GCC} = 
\sqrt{1 - \frac{\langle N_{\rm ch} \rangle}
{\langle N_{\rm ch} \rangle_{\rm tot}}} - 1 \;.
\ee
The formula (\ref{GCC}) derived in \cite{Zaranek:2001di} is
actually approximate as it is derived under the assumption
that the total system's charge $Z$ vanishes. It is, however, 
a reasonable approximation of the exact formula derived in
\cite{Mrowczynski:2001mm} when 
$Z \ll \langle N_{\rm ch} \rangle_{\rm tot}$.

Fig.~10 shows the electric charge fluctuations when the
effect of the global charge conservation is subtracted
that is there is presented 
$\Delta \Phi_q \equiv \Phi_q - \Phi_q^{\rm GCC}$. 
Fig.~10 also shows the levels of charge fluctuations
in the quark-gluon plasma and in the hadronized system,
both computed in a rather simplified model. As seen, the 
observed fluctuations agree very well with the hadron
gas prediction.

\begin{figure}[t]
\centering
\includegraphics*[width=8cm]{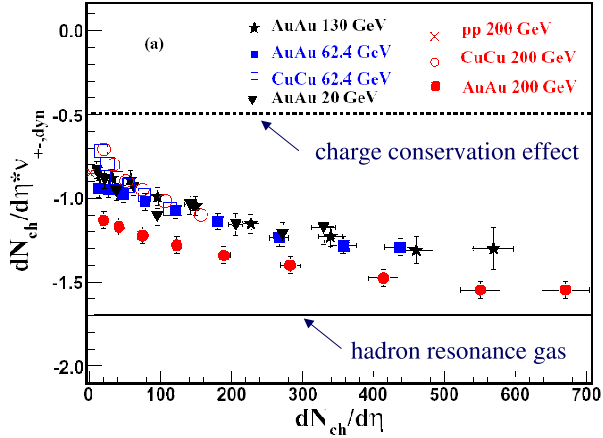}
\caption{Electric charge fluctuations quantified by 
$\nu_{+-}^{\rm dyn}$ as a function of pseudorapidity 
density of charged particles in nucleus-nucleus 
collisions at RHIC collision energies. The vertical axis
shows $\nu_{+-}^{\rm dyn}$ multiplied by the pseudorapidity 
density of charged particles. The figure is taken from 
\cite{Abelev:2008jg}.}
\end{figure}

The electric charge fluctuations were measured at RHIC by 
PHENIX \cite{Adcox:2002mm} and STAR 
\cite{Adams:2003st,Abelev:2008jg} Collaborations. The results
shown in Fig.~11, which is taken from \cite{Abelev:2008jg},
are rather similar to those obtained at SPS. However, the STAR
Collaboration used the measure $\nu_{+-}^{\rm dyn}$ to quantify
the electric charge fluctuations. It is defined as
\be
\nu_{+-}^{\rm dyn} \equiv
 \frac{\langle N_+( N_+ -1) \rangle}{\langle N_+\rangle^2}
+\frac{\langle N_-( N_- -1) \rangle}{\langle N_-\rangle^2}
-2 \frac{\langle N_+ N_-\rangle}
{\langle N_+\rangle \langle N_-\rangle } \;.
\ee
$\nu_{+-}^{\rm dyn}$ is sensitive only to the dynamic 
fluctuations in this sense that it vanishes when the
fluctuations of both $N_+$ and $N_-$ are Poissonian.

As seen in Fig.~11, the observed fluctuations are not only 
bigger than those in QGP but they are even bigger than those 
in the hardon resonance gas. Although we have good reason to 
claim that the quark-gluon plasma is produced at the early 
stage of relativistic heavy-ion collisions at RHIC, the final
state charge fluctuations do not signal the presence of the
QGP. Most probably the fluctuations generated at the plasma
phase are simply washed out during the subsequent system's 
evolution. The fact that the observed charge fluctuations 
are bigger than those in the hardon resonance gas is presumably
caused by a relatively small acceptance of the measurement. 
When a significant fraction of particles originate from 
neutral resonances, which decay into one positive and one 
negative particles, the charge fluctuations are reduced,
when compared to the Poissonian fluctuations, if both 
particles from the decay are observed \cite{Zaranek:2001di}. 
When the experimental acceptance is so small that typically 
only one particle from a resonance decay is registered, the
the electric charge fluctuations remain Poissonian. 

\section{Balance functions}

In the previous section I discussed bulk fluctuations of 
electric charge which at the end appeared to be not very 
informative. Here I am going to present a very interesting 
idea \cite{Bass:2000az,Jeon:2001ue} to measure correlations of 
the electric charges in rapidity by means of the so-called balance 
functions defined as
\be
\label{balance-def}
B(\Delta y) \equiv \frac{1}{2} \bigg[
  \frac{\langle N_{+-}(\Delta y) \rangle 
      - \langle N_{--}(\Delta y) \rangle}
{\langle N_-(\Delta y) \rangle }
+ \frac{\langle N_{-+}(\Delta y) \rangle 
- \langle N_{++}(\Delta y) \rangle }
{\langle N_+(\Delta y) \rangle } \bigg]
\ee
where $\langle N_\pm(\Delta y) \rangle$ and 
$\langle N_{\pm \pm}(\Delta y) \rangle$ are, respectively,
the average number of positive or negative particles 
and the average number of pairs of particles of given
charges within the rapidity (or pseudorapidity) interval 
$\Delta y$ ($\Delta \eta$). The balance functions were argued 
\cite{Bass:2000az,Bialas:2003bb} be sensitive to a hadronization
mechanism. The width of the balance functions was expected
to be bigger, when the hadronization proceeds via the break-up 
of strings as in p-p collisions, than when the quark-gluon 
plasma hadronizes due to the coalescence of constituent quarks. 

\begin{figure}[t]
\begin{minipage}{6cm}
\centering
\vspace{-7mm}
\includegraphics*[width=6cm]{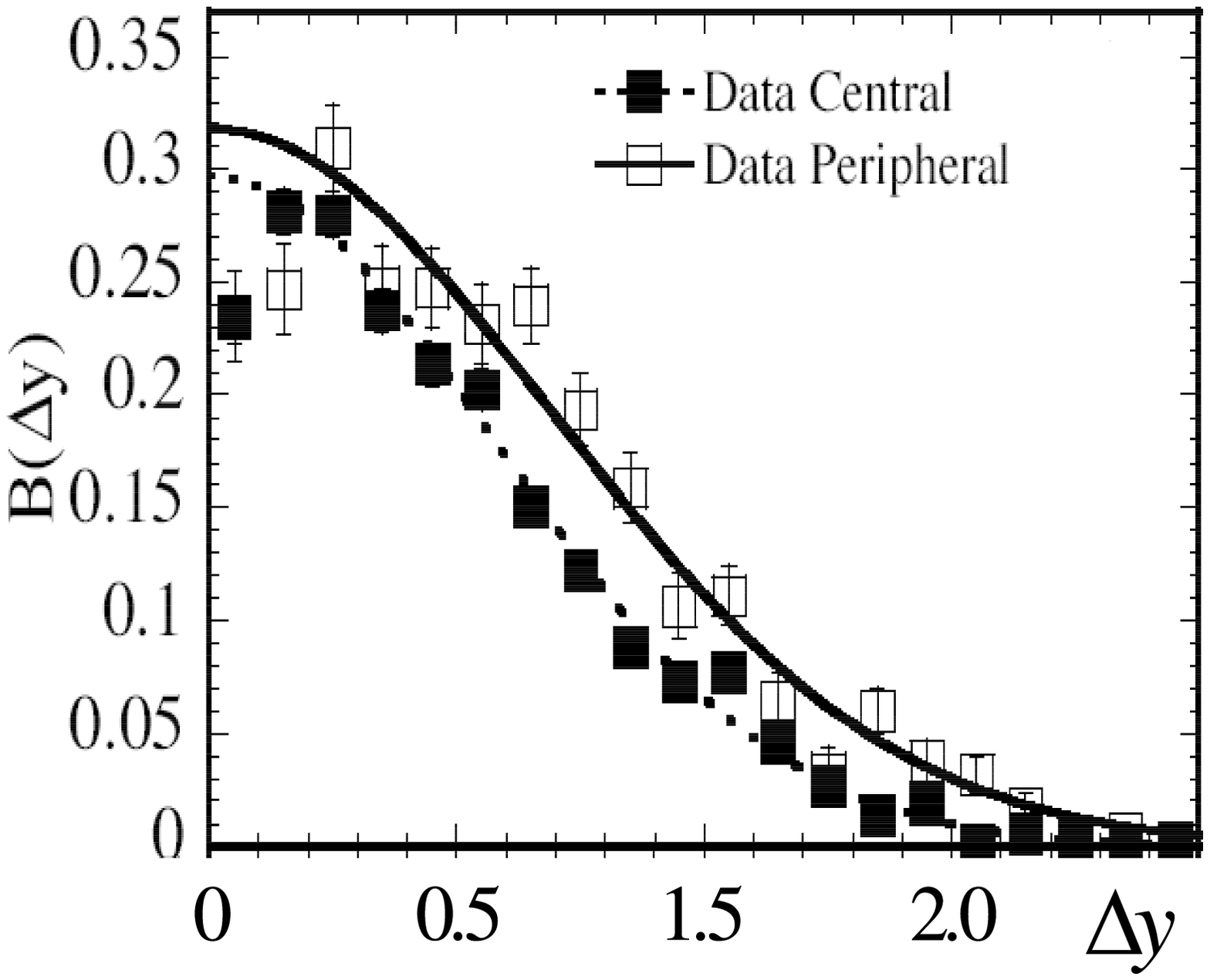}
\caption{The balance functions in central and peripheral
Au-Au collisions $\sqrt{s_{NN}}=130$ GeV. 
The figure is taken from \cite{Adams:2003kg}.}
\end{minipage}
\hspace{3mm}
\begin{minipage}{6cm}
\vspace{-5mm}
\centering
\includegraphics*[width=6cm]{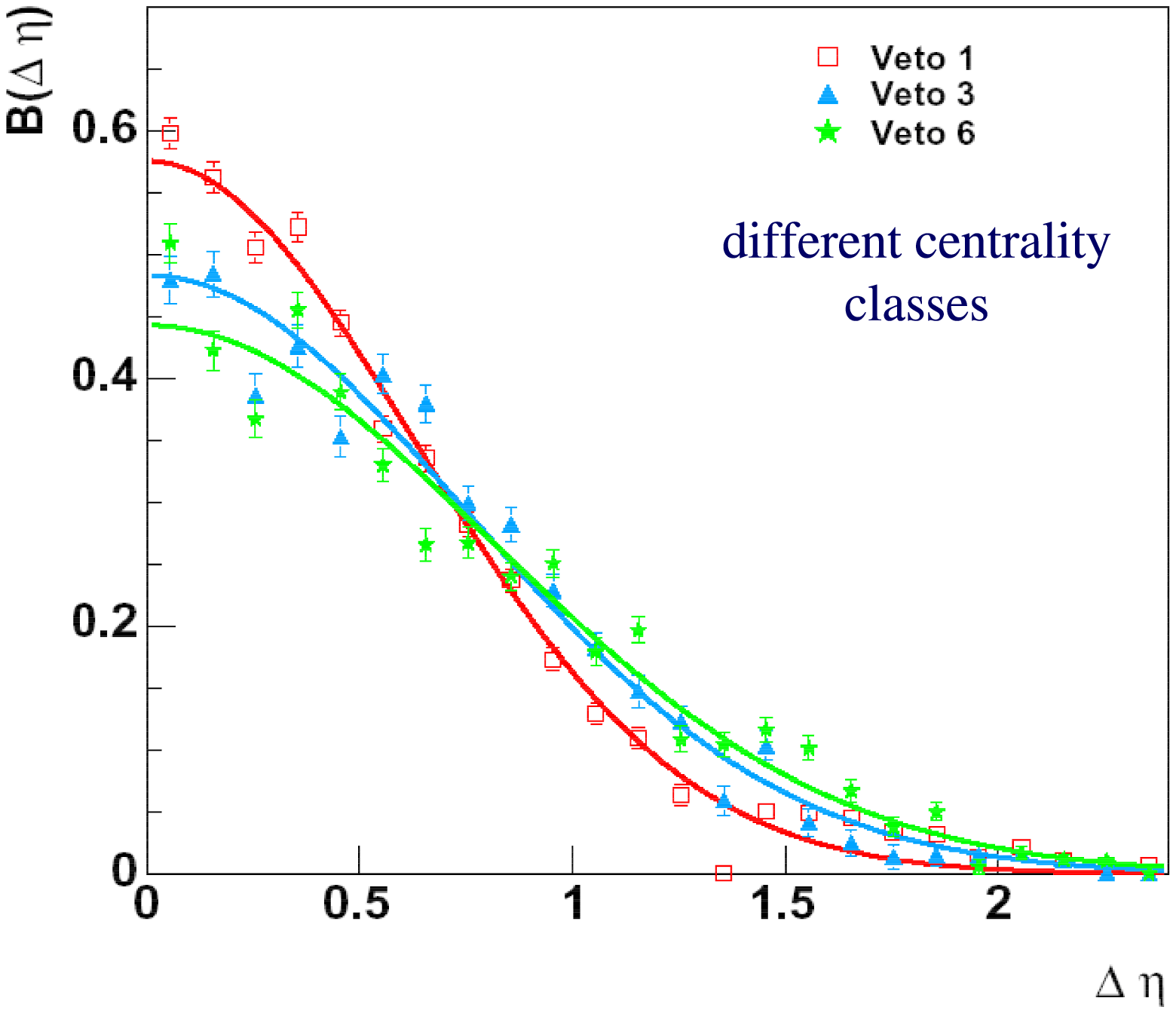}
\caption{The balance functions in Pb-Pb collisions 
at different centralities at 158 AGeV. The centrality 
class `Veto 1' corresponds to the most central 
collisions. The figure is taken from \cite{Alt:2004gx}.}
\end{minipage}
\end{figure}

The balance functions were measured in Au+Au collisions 
at RHIC \cite{Adams:2003kg} and in Pb-Pb collisions at SPS 
\cite{Alt:2004gx}, see Fig.~12 and 13. The balance functions 
for peripheral collisions appeared to have widths consistent 
with model predictions based on a superposition of 
nucleon-nucleon scattering. Widths in central collisions 
were smaller, consistent with trends predicted by models 
incorporating late hadronization due to the coalescence
mechanism. Unfortunately, the interpretation appeared to
be not unique as the balance functions were shown to be
influenced by various factors 
\cite{Pratt:2003gh,Bozek:2003qi,Cheng:2004zy}.
In particular, it was observed that the variation of the 
amount of transverse flow with collision centrality 
can reproduce \cite{Cheng:2004zy} the experimentally observed 
narrowing of the balance functions for central collisions.

\begin{figure}[t]
\centering
\includegraphics*[width=8cm]{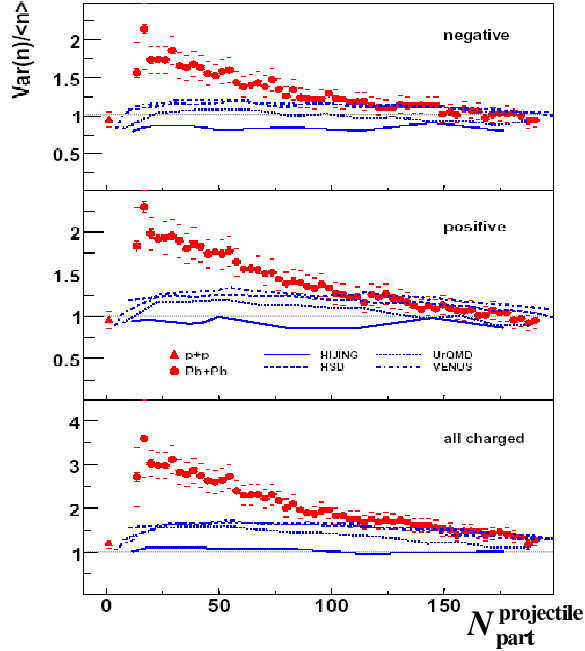}
\caption{The scaled variance of multiplicity distribution 
of negative (upper panel), positive (middle panel) and charged
(lower panel) particles as a function of number of projectile 
participants in nucleus-nucleus collisions at 158 AGeV. The
predictions of HIJING, VENUS, UrQMD and HSD models are also
shown. The figure is taken from \cite{Alt:2006jr}.}
\end{figure}

\section{Multiplicity fluctuations}

As discussed in Sec.~\ref{sec-measure-qunat}, the multiplicity
measurements like that one presented in Fig.~2 are not very
useful, as the results crucially depend on the collision centrality.
The situation is changed if the centrality condition does not result
from specific features of a detector used in the measurement but
if the centrality condition corresponds to a well defined physical
criterion. Such measurements were performed by the NA49 Collaboration
\cite{Alt:2006jr} with the help of zero degree calorimeter which 
allowed one to determine the number of participating nucleons from
a projectile ($N_{\rm part}^{\rm projectile}$) in a given 
nucleus-nucleus collisions. Fig.~14 shows the scaled variance 
($\langle (N - \langle N \rangle )^2 \rangle /\langle N \rangle$)
as a function of $N_{\rm part}^{\rm projectile}$ in p-p and Pb-Pb
collisions at 158 AGeV \cite{Alt:2006jr}. We observe a non-monotonic
behavior of $\langle (N - \langle N \rangle )^2 \rangle /\langle N \rangle$
which contradicts commonly applied models. In the Wounded Nucleon 
Model \cite{Bialas:1976ed}, where produced particles come from wounded 
nucleons, which are assumed to be independent from each other, the 
scale variance is exactly independent of $N_{\rm part}^{\rm projectile}$. 
As seen in Fig.~14, the transport models HIJING \cite{Wang:1991hta}, 
VENUS \cite{Werner:1993uh}, UrQMD \cite{Bass:1998ca} or HSD 
\cite{Cassing:1999es} predict the approximate independence.
It should be noted here that although the scaled variance is
a non-monotonic function of $N_{\rm part}^{\rm projectile}$,
the average multiplicity is simply proportional to 
$N_{\rm part}^{\rm projectile}$ \cite{Alt:2006jr} in agreement 
with the models mentioned above. Although there were several 
theoretical attempts 
\cite{Rybczynski:2004zi,Gazdzicki:2005rr,Cunqueiro:2005hx,Brogueira:2005cn}
to explain the data shown in Fig.~14, in my opinion, there 
is no reliable explanation.

\begin{figure}[t]
\begin{minipage}{6cm}
\vspace{-27mm}
\centering
\includegraphics*[width=6cm]{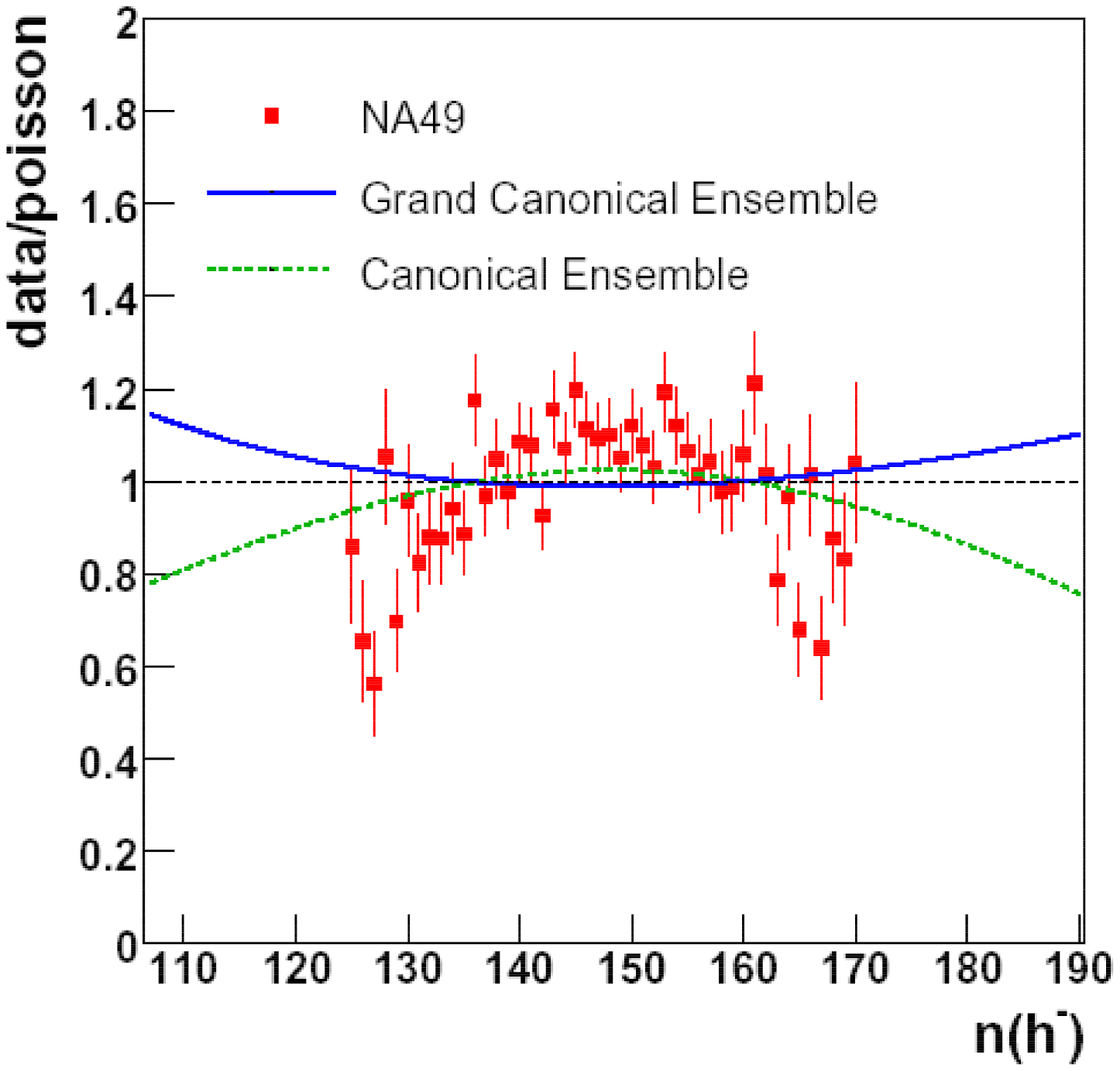}
\caption{The multiplicity distribution of negative charge particles
produced in the most central Pb-Pb collisions at 158 AGeV.
The distribution is divided by the Poisson distribution 
of the same mean. The predictions of statistical models based
on the Grand Canonical and Canonical Ensembles are also shown.
The figure is taken from \cite{Alt:2007jq}.}
\end{minipage}
\hspace{3mm}
\begin{minipage}{6cm}
\vspace{1mm}
\centering
\includegraphics*[width=6cm]{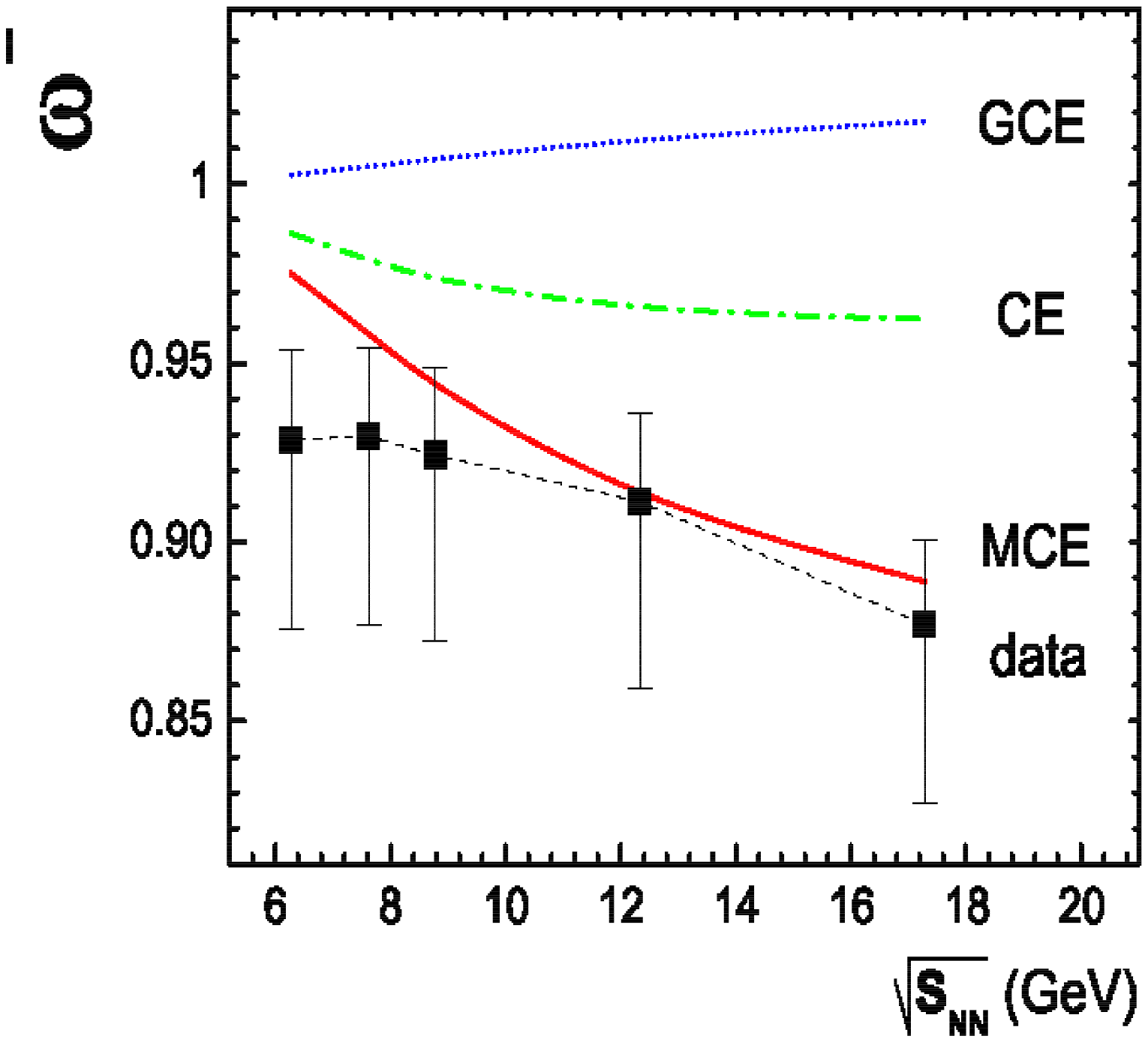}
\caption{The scaled variance of multiplicity distribution 
of negative particles produced in the most central Pb-Pb 
collisions as a function of collision energy. The predictions 
of statistical models based on the Grand Canonical, Canonical 
and Microcanonical Ensembles are also shown. The figure is 
taken from \cite{Begun:2006uu}.}
\end{minipage}
\end{figure}

The multiplicity distribution at the most central collisions
reveals an interesting feature. As shown in Fig.~15 taken form 
\cite{Alt:2007jq} it is narrower not only than the Poisson 
distribution but it is narrower than the multiplicity distribution
obtained in the statistical model \cite{Begun:2006uu} which uses 
the Canonical Ensemble where the electric charge is exactly
conserved. As seen in Fig.~16, this feature persists in a broad 
range of collision energies. Fig.~16 also shows that within
statistical models one has to refer to a microcanonical ensemble
to reproduce the scaled variance of multiplicity distribution.

The multiplicity distributions discussed here appear to be
associated with the transverse momentum fluctuations discussed
in Sec.~\ref{sec-pT-fluc}. As seen in the definition of the
measure $\Phi$ (\ref{Phi-def}), it depends on the multiplicity
distribution. It was shown in \cite{Mrowczynski:2004cg}
that the correlation of the event's transverse momentum and
multiplicity, which is observed in p-p collisions 
\cite{Kafka:1976py},  combined with the non-monotonic scaled 
variance of multiplicity distribution shown in Fig.~14 approximately 
reproduces the $p_T$ fluctuations shown Fig.~4. Therefore, the 
similarity of Figs.~4, 14
is far not superficial.

\section{Elliptic Flow Fluctuations}

The elliptic flow is caused by an azimuthally asymmetric shape 
of the initial interaction zone of colliding nuclei. Consequently,
it is mostly generated in the collision early stage. Fluctuations 
of the elliptic flow were argued to carry information on very early 
stages of relativistic heavy-ion collisions 
\cite{Mrowczynski:2002bw,Mrowczynski:2005gw}. Large fluctuations 
of the elliptic flow were indeed observed at RHIC by PHOBOS 
\cite{Alver:2007rm} and STAR \cite{Sorensen:2006nw} Collaborations.
However, STAR Collaboration claimed later on \cite{Sorensen:2006nw-2} 
that the magnitude of the fluctuations should be taken only as an 
upper limit due to the difficulties to disentangle the elliptic 
flow fluctuations and the contributions which are not correlated 
with the reaction plane. PHOBOS Collaboration has not retracted 
the data \cite{Alver:2007rm}. The whole problem is discussed in 
detail in the very recent review \cite{Voloshin:2008dg}. 

As seen in Fig.~17, the relative $v_2$ fluctuations measured 
by PHOBOS Collaboration \cite{Alver:2007rm} are as large as 
about 40\%. It appears, however, that the effect is dominated not 
by the dynamics but by simple geometrical fluctuations of 
the eccentricity of the interaction zone as suggested in 
\cite{Miller:2003kd}. Since the positions of nucleon-nucleon 
interactions fluctuate within the overlap region of the colliding 
nuclei as illustrated in Fig.~18 taken from \cite{Broniowski:2007ft}, 
the eccentricity of the region fluctuates as well. Since the elliptic
flow is proportional to the eccentricity, the relative eccentricity
fluctuations directly contribute to the relative elliptic flow 
fluctuations. The calculations of the eccentricity fluctuations 
reproduce well the experimentally observed elliptic flow fluctuations, 
see e.g. \cite{Broniowski:2007ft}. Therefore, the hydrodynamic 
evolution of the system, when the elliptic flow is generated, 
seems to be fully deterministic. The result is rather paradoxical 
if one remembers that the elliptic flow is mostly generated at 
a very early stage of the collision when the produced matter 
is presumably not in a complete equilibrium yet.

\begin{figure}[t]
\begin{minipage}{6cm}
\centering
\includegraphics*[width=6cm]{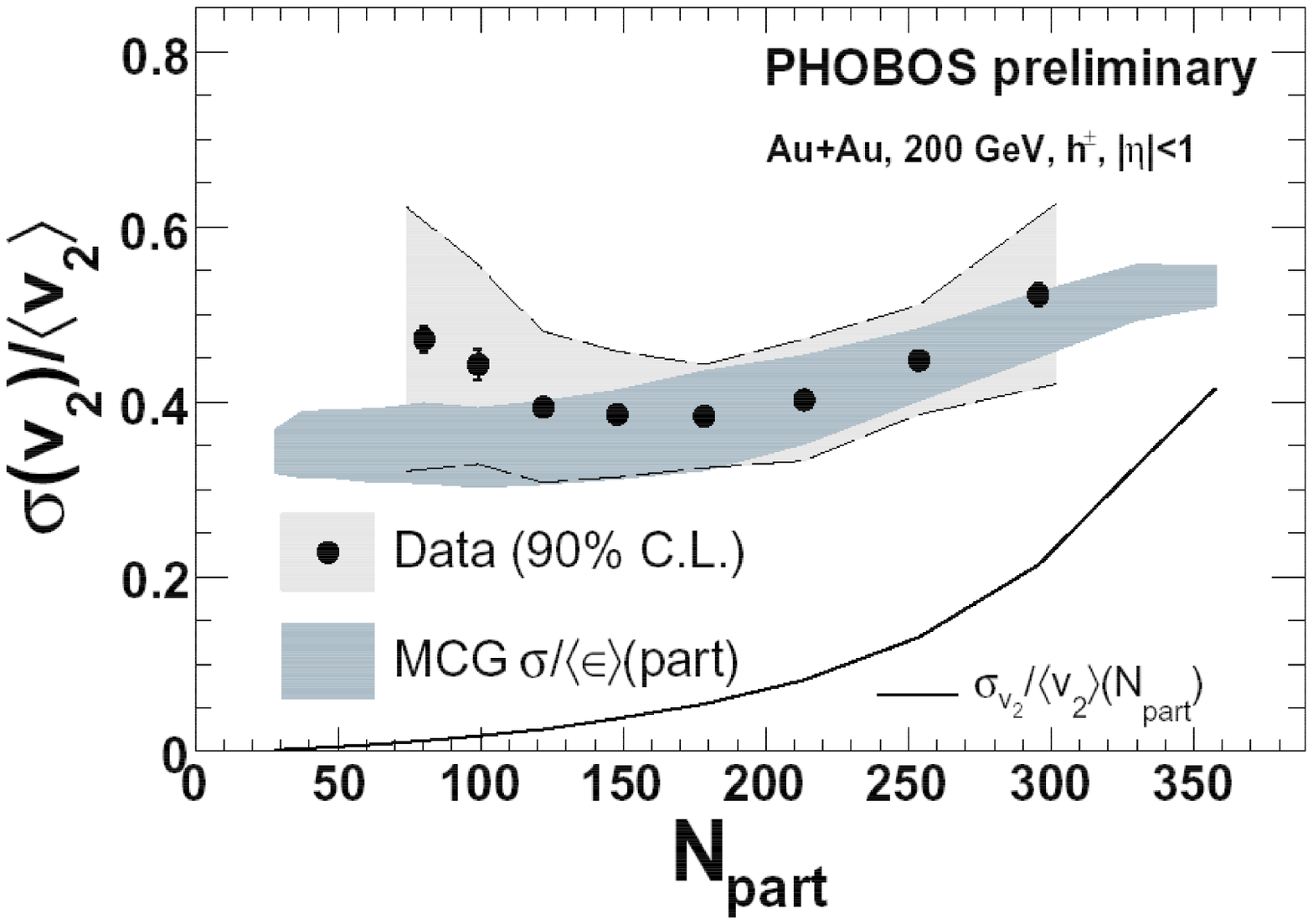}
\caption{The relative fluctuations of the elliptic flow 
in Au-Au collisions at $\sqrt{s_{NN}} = 200$ GeV. 
The figure is taken from \cite{Alver:2007rm}.}
\end{minipage}
\hspace{3mm}
\begin{minipage}{6cm}
\vspace{-7mm}
\centering
\includegraphics*[width=5cm]{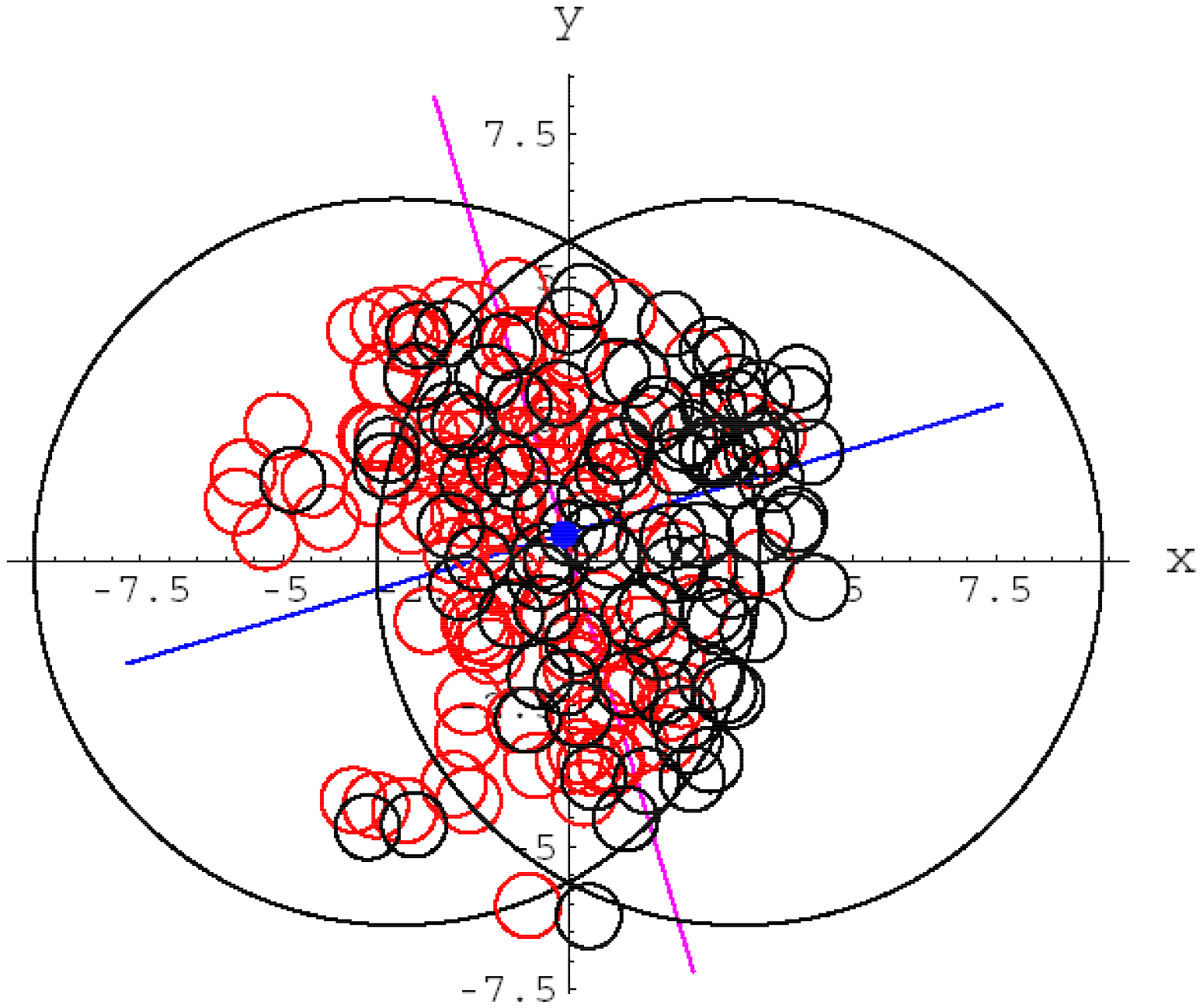}
\vspace{3mm}
\caption{Positions of wounded nucleons in the plane transverse
to the beam in the Au-Au collision. The figure is taken from \cite{Broniowski:2007ft}.}
\end{minipage}
\end{figure}

\section{Conclusions and Outlook}

A big volume of experimental data on event-by-event fluctuations
in relativistic heavy-ion collisions has been collected for last
fifteen years. Some results are indeed very interesting but 
the observed fluctuations are usually dominated by statistical 
noise as convincingly illustrated by similarity of mixed and real 
events. Theoretical expectations of large fluctuations cased by, 
say, phase transitions appeared to be far too optimistic but 
measuring of the fluctuations has also appeared rather difficult.

When single particle distributions are measured a detector 
inefficiency is not a serious obstacle. A number of undetected 
particles should be estimated and the single particle distribution 
is then easily corrected. In the case of correlation measurements, 
the effect of lost particles on the measured correlation depends 
on how the lost particles are correlated with the detected one. Since 
the correlation is {\it a priori} not known, it is unclear how the 
observed correlation should be corrected. For this reason, the 
correlation measurements were usually performed in rather small 
acceptances where the detector efficiency is almost perfect. Then, 
the observed correlation signal does not need a correction for lost 
particles. However, dynamical correlations are usually strongly 
diluted due to a small acceptance. As an example, let me consider 
a multiplicity distribution. If we detect only a small 
fraction $p$ of all particles, the observed multiplicity tends 
to the Poisson distribution when $p \to 0$. Consequently, we 
observe the Poisson distribution in a small acceptance independently 
of the actual distribution. I note that currently no more than 
20\% but typically only a few percent of all produced particles 
are used in event-by-event studies.

Another problem of the current experiments is that the
actual colliding system is not well known as an averaging over 
a centrality interval is performed. Such an averaging dilutes
a potential signal, as most of characteristics of heavy-ion
collisions strongly depends on centrality. Sometimes the centrality 
is estimated using produced particles which are analyzed. Then, 
the effect of autocorrelation has to be additionally removed 
from the data. 

The analysis of multiplicity clearly shows how important is
a good determination of centrality. The multiplicity 
measurement presented in Fig.~2 badly depends on experimental 
condition and thus is not very useful. When the collision
centrality is so precisely measured that the number of participating
nucleons from a projectile is known, the multiplicity distribution
appeared to conceal very interesting features displayed in
Fig.~14, 15.

As the observed dynamical fluctuations are usually small, it is 
difficult to extract physically interesting information, it is 
even more difficult to workout a unique interpretation. New 
theoretical ideas and reliable models are certainly needed but 
what the event-by-event physics really requires is, in my opinion, 
a new generation of experiments which will fulfill two important 
conditions: i) the acceptance is a sizeable fraction of $4\pi$,
ii) the collision centrality is measured up to single nucleons
participating in a collision. The future NA61/SHINE program at 
SPS is hoped to satisfy the requirements \cite{Gazdzicki:2008kk}.

\vspace{5mm} 

This work was partially supported by Polish Ministry of Science 
and Higher Education under grant N N202 3956 33.



\begin{thebibliography}{99}

\bibitem{Heiselberg:2000fk}
H.~Heiselberg,
Phys.\ Rept.\  {\bf 351}, 161 (2001).

\bibitem{Jeon:2003gk}
S.~Jeon and V.~Koch,
in {\it Quark Gluon Plasma 3}, edited by R.C. Hwa and X.-N. Wang
(Scientific, Singapore, 2004).

\bibitem{Stock:1994ve}
R.~Stock,
in {\it Proceedings of NATO Advanced Study Workshop on Hot Hadronic 
Matter: Theory and Experiment}, Divonne-les-Bains, France, 
June 27 - July 1, 1994, edited by J. Letessier, H. H. Gutbrod, 
and J. Rafelski (Plenum Press, New York, 1995).

\bibitem{Appelshauser:1999ft}
H.~Appelshauser {\it et al.}  [NA49 Collaboration],
Phys.\ Lett.\  B {\bf 459}, 679 (1999).

\bibitem{Aggarwal:2001aa}
M.~M.~Aggarwal {\it et al.}  [WA98 Collaboration],
Phys.\ Rev.\  C {\bf 65}, 054912 (2002).

\bibitem{Adcox:2002pa}
K.~Adcox {\it et al.}  [PHENIX Collaboration],
Phys.\ Rev.\  C {\bf 66}, 024901 (2002).

\bibitem{Gazdzicki:1992ri}
M.~Ga\'zdzicki and St.~Mr\'owczy\'nski,
Z.\ Phys.\  C {\bf 54}, 127 (1992).

\bibitem{Bialas:1976ed}
A.~Bia\l as, M.~B\l eszynski and W.~Czy\.z,
Nucl.\ Phys.\  B {\bf 111}, 461 (1976).

\bibitem{Wang:1991hta}
X.~N.~Wang and M.~Gyulassy,
Phys.\ Rev.\  D {\bf 44}, 3501 (1991).

\bibitem{Werner:1993uh}
K.~Werner,
Phys.\ Rept.\  {\bf 232}, 87 (1993).
 
\bibitem{Bass:1998ca}
S.~A.~Bass {\it et al.},
Prog.\ Part.\ Nucl.\ Phys.\  {\bf 41}, 255 (1998).

\bibitem{Cassing:1999es}
W.~Cassing and E.~L.~Bratkovskaya,
Phys.\ Rept.\  {\bf 308}, 65 (1999).
\bibitem{Voloshin:1999yf}
S.~A.~Voloshin, V.~Koch and H.~G.~Ritter,
Phys.\ Rev.\  C {\bf 60}, 024901 (1999).

\bibitem{Trainor:2000dm}
T.~A.~Trainor,
arXiv:hep-ph/0001148.

\bibitem{Adamova:2003pz}
D.~Adamova {\it et al.}  [CERES Collaboration],
Nucl.\ Phys.\  A {\bf 727}, 97 (2003).

\bibitem{Adler:2003xq}
S.~S.~Adler {\it et al.}  [PHENIX Collaboration],
Phys.\ Rev.\ Lett.\  {\bf 93}, 092301 (2004).

\bibitem{Anticic:2003fd}
T.~Anticic {\it et al.}  [NA49 Collaboration],
Phys.\ Rev.\  C {\bf 70}, 034902 (2004).

\bibitem{Bialas:1990dk}
A.~Bia\l as and M.~Ga\'zdzicki,
Phys.\ Lett.\  B {\bf 252}, 483 (1990).

\bibitem{Grebieszkow:2007xz}
K.~Grebieszkow {\it et al.} [NA49 Collaboration],
PoS {\bf CPOD07}, 022 (2007).

\bibitem{Adamova:2008sx}
D.~Adamova {\it et al.} [CERES Collaboration],
Nucl.\ Phys.\  A {\bf 811}, 179 (2008).

\bibitem{Kafka:1976py}
T.~Kafka {\it et al.},
Phys.\ Rev.\  D {\bf 16}, 1261 (1977).

\bibitem{Mrowczynski:2004cg}
St.~Mr\'owczy\'nski, M.~Rybczy\'nski and Z.~W\l odarczyk,
Phys.\ Rev.\ C {\bf 70}, 054906 (2004).

\bibitem{Adams:2003uw}
J.~Adams {\it et al.} [STAR Collaboration],
Phys.\ Rev.\  C {\bf 71}, 064906 (2005).

\bibitem{Broniowski:2006zz}
W.~Broniowski, P.~Bo\.zek, W.~Florkowski and B.~Hiller,
PoS {\bf CFRNC2006}, 020 (2006).

\bibitem{Lan-Lif}
L.D.~Landau and E.M.~Lifshitz, {\it Statistical Physics}
(Pergamon Press, Oxford, 1980). 

\bibitem{Stodolsky:1995ds}
L.~Stodolsky,
Phys.\ Rev.\ Lett.\  {\bf 75}, 1044 (1995).

\bibitem{Shuryak:1997yj}
E.~V.~Shuryak,
Phys.\ Lett.\  B {\bf 423}, 9 (1998).

\bibitem{Kit88} C.~Kittel, Phys. Today {\bf 41}, 93 (1988).

\bibitem{Man88} B.~Mandelbrot, Phys. Today {\bf 42}, 71 (1988).

\bibitem{Stephanov:1999zu}
M.~A.~Stephanov, K.~Rajagopal and E.~V.~Shuryak,
Phys.\ Rev.\  D {\bf 60}, 114028 (1999).

\bibitem{Mrowczynski:1997kz}
St.~Mr\'owczy\'nski,
Phys.\ Lett.\  B {\bf 430}, 9 (1998).

\bibitem{Jeon:2000wg}
S.~Jeon and V.~Koch,
Phys.\ Rev.\ Lett.\  {\bf 85}, 2076 (2000).

\bibitem{Asakawa:2000wh}
M.~Asakawa, U.~W.~Heinz and B.~Muller,
Phys.\ Rev.\ Lett.\  {\bf 85}, 2072 (2000).

\bibitem{Alt:2004ir}
C.~Alt {\it et al.}  [NA49 Collaboration],
Phys.\ Rev.\  C {\bf 70}, 064903 (2004).

\bibitem{Zaranek:2001di}
J.~Zaranek,
Phys.\ Rev.\  C {\bf 66}, 024905 (2002).

\bibitem{Mrowczynski:2001mm}
St.~Mr\'owczy\'nski,
Phys.\ Rev.\  C {\bf 66}, 024904 (2002).

\bibitem{Adcox:2002mm}
K.~Adcox {\it et al.}  [PHENIX Collaboration],
Phys.\ Rev.\ Lett.\  {\bf 89}, 082301 (2002).

\bibitem{Adams:2003st}
J.~Adams {\it et al.}  [STAR Collaboration],
Phys.\ Rev.\  C {\bf 68}, 044905 (2003).

\bibitem{Abelev:2008jg}
B.~I.~Abelev {\it et al.}  [STAR Collaboration],
arXiv:0807.3269 [nucl-ex].

\bibitem{Bass:2000az}
S.~A.~Bass, P.~Danielewicz and S.~Pratt,
Phys.\ Rev.\ Lett.\  {\bf 85}, 2689 (2000)

\bibitem{Jeon:2001ue}
S.~Jeon and S.~Pratt,
Phys.\ Rev.\  C {\bf 65}, 044902 (2002).
\bibitem{Bialas:2003bb}
A.~Bia\l as,
Phys.\ Lett.\  B {\bf 579}, 31 (2004).

\bibitem{Adams:2003kg}
J.~Adams {\it et al.}  [STAR Collaboration],
Phys.\ Rev.\ Lett.\  {\bf 90}, 172301 (2003).

\bibitem{Alt:2004gx}
C.~Alt {\it et al.}  [NA49 Collaboration],
Phys.\ Rev.\  C {\bf 71}, 034903 (2005).

\bibitem{Pratt:2003gh}
S.~Pratt and S.~Cheng,
Phys.\ Rev.\  C {\bf 68}, 014907 (2003).

\bibitem{Bozek:2003qi}
P.~Bo\.zek, W.~Broniowski and W.~Florkowski,
Acta Phys.\ Hung.\  A {\bf 22}, 149 (2005).

\bibitem{Cheng:2004zy}
S.~Cheng {\it et al.},
Phys.\ Rev.\  C {\bf 69}, 054906 (2004).

\bibitem{Alt:2006jr}
C.~Alt {\it et al.}  [NA49 Collaboration],
Phys.\ Rev.\  C {\bf 75}, 064904 (2007).

\bibitem{Rybczynski:2004zi}
M.~Rybczy\'nski and Z.~W\l odarczyk,
J.\ Phys.\ Conf.\ Ser.\  {\bf 5}, 238 (2005).

\bibitem{Gazdzicki:2005rr}
M.~Ga\'zdzicki and M.~I.~Gorenstein,
Phys.\ Lett.\  B {\bf 640}, 155 (2006).

\bibitem{Cunqueiro:2005hx}
L.~Cunqueiro, E.~G.~Ferreiro, F.~del Moral and C.~Pajares,
Phys.\ Rev.\ C {\bf 72}, 024907 (2005).

\bibitem{Brogueira:2005cn}
P.~Brogueira and J.~Dias de Deus,
Phys.\ Rev.\ C {\bf 72}, 044903 (2005).

\bibitem{Alt:2007jq}
C.~Alt {\it et al.}  [NA49 Collaboration],
arXiv:0712.3216 [nucl-ex].

\bibitem{Begun:2006uu}
V.~V.~Begun, M.~Ga\'zdzicki, M.~I.~Gorenstein, M.~Hauer, V.~P.~Konchakovski and B.~Lungwitz,
Phys.\ Rev.\  C {\bf 76}, 024902 (2007).

\bibitem{Mrowczynski:2002bw}
St.~Mr\'owczy\'nski and E.~V.~Shuryak,
Acta Phys.\ Polon.\  B {\bf 34}, 4241 (2003).
 
\bibitem{Mrowczynski:2005gw}
St.~Mr\'owczy\'nski,
J.\ Phys.\ Conf.\ Ser.\  {\bf 27}, 204 (2005).

\bibitem{Alver:2007rm}
B.~Alver {\it et al.}  [PHOBOS Collaboration],
J.\ Phys.\ G {\bf 34}, S907 (2007).

\bibitem{Sorensen:2006nw}
P.~Sorensen [STAR Collaboration],
J.\ Phys.\ G {\bf 34}, S897 (2007).

\bibitem{Sorensen:2006nw-2}
P.~Sorensen  [STAR Collaboration],
[arXiv:nucl-ex/0612021].

\bibitem{Voloshin:2008dg}
S.~A.~Voloshin, A.~M.~Poskanzer and R.~Snellings,
arXiv:0809.2949 [nucl-ex].

\bibitem{Miller:2003kd}
M.~Miller and R.~Snellings,
arXiv:nucl-ex/0312008.

\bibitem{Broniowski:2007ft}
W.~Broniowski, P.~Bo\.zek and M.~Rybczy\'nski,
Phys.\ Rev.\  C {\bf 76}, 054905 (2007).

\bibitem{Gazdzicki:2008kk}
M.~Ga\'zdzicki [NA61/SHINE Collaboration],
arXiv:0812.4415 [nucl-ex].

\end{thebibliography}
\end{document}